\documentclass[a4paper,11pt,final]{article}

\usepackage{graphicx,amsmath,amssymb} 
\usepackage{epsfig}

\begin{document}

\title{\ \\ \LARGE\bf  Analyzing coevolutionary games with dynamic fitness landscapes}

\author{Hendrik Richter \\
HTWK Leipzig University of Applied Sciences \\ Faculty of
Electrical Engineering and Information Technology\\
        Postfach 301166, D--04251 Leipzig, Germany. \\ Email: 
hendrik.richter@htwk-leipzig.de. }

\maketitle

\begin{abstract}
Coevolutionary games cast players that may change their strategies as well as their networks of interaction. In this paper a framework is introduced for describing coevolutionary game dynamics by landscape models. It is shown that coevolutionary games invoke dynamic landscapes. Numerical experiments are shown for a prisoner's dilemma (PD) and a snow drift (SD) game that both use either birth--death (BD) or death--birth (DB) strategy updating. The resulting landscapes are analyzed with respect to modality and  ruggedness.
\end{abstract}

\section{Introduction}
Evolutionary games are mathematical models of  interactions between individuals in a population and explain how  their behavioral strategies (for instance cooperation or competition)  contribute to payoff collecting and consequently to the individuals' success in terms of fitness.  An evolutionary game becomes dynamic if it is played iteratively over several rounds and the individuals are allowed to change their strategies and/or to recast the network describing with whom they are  interacting.  
Such an iterated evolutionary game comprises of an evolving population of individuals acting as players and can be seen as an expression of evolutionary dynamics.  

For describing evolutionary dynamics the  framework of fitness landscapes has been introduced, e.g.~\cite{kauffm91,richengel14,stad03}.   A fitness landscape formulates relationships between the genetic specifications of  individuals and their fitness. Together with the postulate of differences in fitness over all possible genetic specifications and a moving bias towards higher fitness, the setup suggests the picture of an evolving  population that is moving directedly on the landscape. On a conceptual level, this picture is based on the notion of evolutionary paths that are created by the topological features of the fitness landscape. Evolutionary paths are a succession of moves on the landscape with   persistently ascending fitness values. The existence and abundance of such evolutionary paths gives raise to estimates about how likely a transition from low--fitness regions to high--fitness regions in the landscape is. These transitions instantiate evolutionary dynamics.  

Given the fact  that there are two frameworks for addressing evolutionary dynamics, it appears naturally to ask about their relationships. Unfortunately, both frameworks are not immediately compatible.
Although it is acknowledged that evolutionary games cast fitness landscapes, it became also clear that such game landscapes change with an evolving population of players~\cite{nowak04}. This is attributed to frequency--dependent selection. In other words, game landscapes are dynamic.  Based on some recent results on dynamic fitness landscapes, e.g.~\cite{foster13,rich08,
rich14b}, there are some first attempts  applying these ideas to games~\cite{rich15}. In this paper the concept of dynamic landscapes is used for analyzing evolutionary games. Games are considered where the players may update their strategies (evolutionary games~\cite{allen14,green12,szabo07}) but also games where the players may additionally change their network of interaction (coevolutionary games~\cite{perc10,tani07}). In particular, it is shown that
the proposed method makes it possible to model and analyze evolutionary games and coevolutionary games within the same framework.
The paper is structured as follows. In Sec.  \ref{sec:desc}, some basic definitions are given and evolutionary and coevolutionary games are briefly recalled. Sec.  \ref{sec:gamedynamic} reviews game dynamics, particularly the processes to update strategies and networks of interaction. Landscape models of coevolutionary games are introduced and discussed in  Sec. \ref{sec:land}.  Sec.  \ref{sec:num} reports numerical  experiments on landscape properties such as modality and ruggedness. Sec. \ref{sec:con} closes the paper with a summary and conclusions.

\section{Definitions and game description}  \label{sec:desc}
The coevolutionary games considered in the paper comprise of three levels of activity: (i) game playing, (iii) updating  the strategy,  and (iii) updating  the network of interaction.  The game playing is done by a population of $N$  players $\mathcal{I}$ that use one of two strategies $\pi$ during each round $k=\{0,1,2,\ldots \}$. A player $\mathcal{I}_i \in \mathcal{I}$, $i=1,2,\ldots,N$, can either cooperate ($C_i$) or defect ($D_i$). A pairwise interaction between two players $\mathcal{I}_i$ and $\mathcal{I}_j$ (which can be seen as player and coplayer) yields a reward in form of a payoff $(p_i,p_j)$ as given by the payoff matrix      
\begin{equation} 
\bordermatrix{~ & C_j & D_j \cr
                  C_i & R & S \cr
                  D_i & T & P \cr}. \label{eq:payoff}
\end{equation}
For player and coplayer using the same strategy, $(\pi_i,\pi_j)=(C_i,C_j)$ or  $(\pi_i,\pi_j)=(D_i,D_j)$,   the reward for mutual cooperation $p_i=p_j=R$ or the punishment for mutual defection $p_i=p_j=P$ is obtained. A mixed choice of strategy  gives   the sucker payoff $S$ for cooperating with a defector, and  the temptation $T$ to defect while the coplayer is cooperating. Hence, for $(\pi_i,\pi_j)=(C_i,D_j)$, there is $p_i=S$ and $p_j=T$, while for $(\pi_i,\pi_j)=(D_i,C_j)$, there is $p_i=T$ and $p_j=S$.  Depending on  the numerical values of $(R,P,S,T)$ and their order, particular examples of the game are obtained, which have become metaphors for studying social dilemmas and discussing strategy selection along with the effect on short-- and long--term success in accumulating payoff~\cite{maysmit91,nowak06}. Most prominently, there are prisoner's dilemma (PD) games, where $T>R>P>S$, and  snow drift (SD) games, where $T>R>S>P$.

The payoff $p_i(k)$ of player $\mathcal{I}_i$ in round $k$ depends not only on  the player's  strategy $\pi_i(k)$ and the values of the payoff matrix (\ref{eq:payoff}), but also on who its coplayer is (or more precisely as to what the coplayer's strategy is) and how many coplayers there are. The question of who--plays--whom in a given round of the game is addressed by the network of interaction. A convenient way of expressing and visualizing the network of interaction is by using elements from evolutionary graph theory~\cite{allen14,lieb05,ohts07,sha12}. Evolutionary graph theory places each player of the population on a vertex of an (undirected) graph. This graph describes the network of interaction and consequently it is called interaction graph.  As there are no empty vertices and a vertex can only be occupied by one player, the number of vertices of the graph equals the number of players $N$. For each player, its coplayers are indicated by edges that connect the vertex of the player with the vertices of the coplayers. Such an edge defines the connected players to be adjacent. Each vertex can have up to $N-1$ edges (self--play is not allowed). As the degree $d$ is the number of edges incident with a vertex, the degrees of the interaction graph equal the number of coplayers that are engaged with each player in a single round. A graph is called regular if the degree is the same for all vertices. Hence, a regular interaction graph means that all players have the same number of coplayers.   

The interaction graph can be described algebraically by its (interaction) adjacency matrix $A_I$, which is also called interaction matrix.  The matrix $A_I \in [0,1]^{N \times N}$ is a symmetric $N \times N$ matrix with  entries $a_{ij}=1$ indicating an edge between vertex $i$ and $j$ and $a_{ij}=0$  showing that players $\mathcal{I}_i$ and $\mathcal{I}_j$ are not adjacent. The diagonal elements $a_{ii}=0$ because there is no self--play.    The symmetry of $A_I$ reflects the fact that two players $\mathcal{I}_i$ and $\mathcal{I}_j$ mutually engage in the game.  From the perspective of player $\mathcal{I}_i$, the player $\mathcal{I}_j$ may be the coplayer. If so, then from the perspective of player $\mathcal{I}_j$, the player $\mathcal{I}_i$ is the coplayer. Consequently, $a_{ij}=a_{ji}$ in the adjacency matrix $A_I$. If all $a_{ij}=1$ (except $a_{ii}=0$), the graph is complete, meaning that every player has all other players as coplayers and the evolutionary game is understood to be well--mixed~\cite{szabo07}.   To summarize, for describing completely and deterministically  the game and the allocation of payoff, apart from the payoff matrix  (\ref{eq:payoff}) only two other entities are necessary: the strategy vector $\pi=(\pi_1 \pi_2 \ldots \pi_N)$ with $\pi_i \in [C_i, D_i]$ and the adjacency matrix $A_I$. This setting deterministically fixes the payoff $p=(p_1,p_2,\ldots,p_N)$ for each player.
 For making payoff $p_i$  of a player $\mathcal{I}_i$  interpretable as reproduction rate or survival probability (and lastly as fitness $f$), it has been suggested to rescale $p_i$ by a positive, increasing, differentiable function~\cite{allen14,sha12}. In the following the linear function $f=1+\delta \cdot p$ is used with  the intensity of selection $\delta\geq 0$. 

\section{Coevolutionary game dynamics} \label{sec:gamedynamic}

As the game is completely determined  by fixing  the payoff matrix  (\ref{eq:payoff}) as well as the strategy vector $\pi$, and the adjacency matrix $A_I$, the distribution of payoff $p_i(k)$ amongst the players remains the same if the
players  were to engage in the game with the same entities for a second
time in round $k+1$. In other words, for these entities being constant the game  can be seen as static.  
Consequently, to make the evolutionary game dynamic  requires to update either the players'
strategies or the network of interaction, or both.

\subsection{Updating Strategies}  \label{sec:statupdate}
There is a huge amount of work devoted to the modes of updating the player strategies in evolutionary games~\cite{allen14,ohts07,patt15}. Most models use versions of a stochastic strategy updating based on a Moran process, but there are also works emphasizing on limiting the effect of random and  including self--interest of players, e.g.~\cite{green14}.  According to a Moran process~\cite{whig08}, in each round a player $\mathcal{I}_i$ (or more precisely its strategy) is replaced by (the strategy of) a player $\mathcal{I}_j$. The players $\mathcal{I}_i$ and $\mathcal{I}_j$ are selected at random, but the probabilities of the selection may not be uniform, for instance  depending on the players' fitness, which may vary. Versions of stochastic updating rules differ in several respects. Differences are, for example,  the actual probabilities that given players $\mathcal{I}_i$ and $\mathcal{I}_j$ are selected 
or whether or not 
there is an order between selecting the player providing the strategy (the source) and selecting the player receiving the strategy (the target).  
Finally, there may be general restrictions as to which players are allowed to be a possible source and/or target of another player.  Such predetermined restrictions imply a replacement structure~\cite{ohts07}.  Conceptual similar to interaction,  the question of who--may--replace--whom can be described by a network of replacement. This network is expressible by a replacement graph and consequently by a (replacement) adjacency matrix $W_R$, which is also called replacement matrix. The 
 matrix $W_R \in \mathbb{R}^{N \times N}$ has  entries $w_{ij}>0$ if player $\mathcal{I}_i$ may provide its strategy for player $\mathcal{I}_j$ to receive.   Note that the values of $w_{ij}>0$ contribute to the probabilities that player $\mathcal{I}_i$ is source and player $j$ is target.   Hence,  if for a constant $\bar{w}\neq0$ all $w_{ij}=\bar{w}$,  every player $\mathcal{I}_i$ may be the source to every target player $\mathcal{I}_j$ with equal probability. In other words, if there are no restrictions, the replacement graph is fully connected with evenly weighted  edges. 

 Amongst  strategy updating, the following replacement rules are frequently studied: birth--death (BD), death--birth (DB),  imitation (IM), and pair--wise comparison (PD)~\cite{allen14,patt15,sha12}. For all rules there may be restrictions with respect to  replacement. The rules BD and DB differ in the order with which source and target are selected, with BD selecting source before target and DB target before source. The probability to become a source depends on the source's fitness.  IM  is similar to DB but with the difference that the target itself can compete with other players to become a source. In PD (also known as link dynamics) both players are selected simultaneously and the source is replaced by the target with a probability depending on  the  fitness difference between the players, for instance via a Fermi function.  Hence, IM and PD share that the target can be its own source, meaning than the strategy remains the same. To summarize,  all Moran--based updating rules
depend only on random (and possibly on players' fitness and replacement restrictions), but not on details of the interaction  (for instance who the source or target are actually interacting with and what those strategies are). Therefore, they do not account for self--interested players~\cite{green14} and make  it possible to disentangle player and strategy in the sense that it makes no difference from which source the target receives its strategy updating. 
In other words, for all these updating rules it is possible to specify probabilities $\tau_{ij}$ that the strategy of a source $\mathcal{I}_i$ replaces the strategy of a target $\mathcal{I}_j$ depending only on replacement matrix and fitness~\cite{patt15}. 

\subsection{Updating networks of interaction} \label{sec:netwupdate}

If in addition to the strategies also the network of interaction can be updated in evolutionary games, the game is called coevolutionary. In essence, coevolution in evolutionary games is to consider the network of interaction as dynamic, from which follows that the interaction matrix $A_I$ must be time--dependent. There is a substantial variety of schemes and rules for coevolution~\cite{pach06,perc10,tani07}. These schemes can be categorized according to different criteria. A first criterion is the type of dynamics of $A_I$, for which there can be three groups: (i) purely random updating, (ii) random updating with probabilities depending on fitness or current strategy or network properties, and (iii) deterministic updating. A second criterion  is the effect which the dynamics has on graph--theoretical properties of the networks, for instance, the number of edges (is the number of links in the network constant or growing/shrinking), or the regularity of the graph (do all players have always the same number of coplayers, or are there rules that allow specific players to become super--connected), or network connectivity.
Finally,  there is the question of time scale,  that is how the cycles of strategy updating relate to the cycles of network updating, for instance if the edges have a life--time depending on the number of strategy updating that the players experienced.  

Unfortunately, the topic of network updating is not yet matured as far as
to express for a given rule transitions from one network to another probabilistically. Whereas for strategy updating, there are replacement probabilities for different updating rules~\cite{patt15}, the same is not known for network updating. 
However, it might be reasonable to assume that a network updating involves to create an interaction matrix $A_I(k+1)$ at point in time $k+1$ from a matrix $A_I(k)$ at the previous point $k$. 
Such a succession of interaction networks can be modelled by instances of an Erd{\"o}s--R{\'e}nyi graph. In this paper, the discussion is restricted to the case that the number of coplayers is the same for all players and constant for all updating instances. Employing such a model precludes situations where a more highly connected player receives high fitness  due to its connectedness, but not necessarily due to the effectiveness of its strategy. For $d$ coplayers, such an interaction graph has degree $d$ and belongs to a special class of Erd{\"o}s--R{\'e}nyi graphs, namely  random $d$--regular graphs. Modelling the interaction network by  random $d$--regular graphs makes it possible to systematically carry out  numerical experiments because recently efficient algorithms for generating such graphs became available~\cite{bay10,blitz11,kim06}. Moreover, for  random $d$--regular graphs, some analytic results about the number of different graphs are known. This, in turn, corresponds to the number of possible player--coplayer combinations. 

For a small number of edges ($=$ coplayers)  $d$, the number $\mathcal{L}_d(N)$ of different   $d$--regular graphs for $N$ vertices ($=$ players) can be found by enumeration, see for instance the entries in the Sloane
encyclopedia of integer sequences~\cite{slon15}. Thus, $\mathcal{L}_2(4)=3$ and $\mathcal{L}_3(4)=1$, while $\mathcal{L}_2(6)=70$,  $\mathcal{L}_3(6)=70$,  $\mathcal{L}_4(6)=15$ and  $\mathcal{L}_5(6)=1$, and $\mathcal{L}_2(8)=3507$, $\mathcal{L}_2(10)=286884$. Note that generally $\mathcal{L}_{N-1}(N)=1$, which means that a well--mixed population represented by a complete network of interaction possesses only one instance of the matrix $A_I$.   
In other words, for a complete network graph the game cannot be coevolutionary. It  is always static with respect to interaction because no dynamic changes can be cast out of a single instance of $A_I$.  
 Further note that $\mathcal{L}_d(N)$ grows rapidly. Although no simple closed formula for $\mathcal{L}_d(N)$ and given $N$ and $d$ is known, asymptotic expressions have been found~\cite{worm99}. Asymptotically, and for $d=\hbox{o}(\sqrt{N})$ and $dN$ even, the number is   \begin{equation} \mathcal{L}_d(N)=\frac{ (dN)! \: \cdot \:  \exp{\left(\frac{1-d^2}{4}-\frac{d^3}{12N}+\mathcal{O}\left (\frac{d^2}{N}\right )\right)}   }{\left(\frac{dN}{2}\right)!\: 2^{\frac{dN}{2}} \: (d!)^N}. \label{eq:LDN} \end{equation}
Based on a collection of  random $d$--regular graphs the effect of different networks of interaction on payoff collecting and fitness can be analyzed, for which a landscape approach is proposed in the next section.    

\section{Landscape models  
of 
game dynamics
} \label{sec:land}

\subsection{Static and dynamic landscapes}

A general definition of a (static) fitness landscape $\Lambda_S$ is the triple $\Lambda_S=(\mathbb{X},n,f)$,
where $\mathbb{X}$ is a configuration space, $n(x)$ is a
neighborhood structure that assigns to every $x \in \mathbb{X}$ a
set of direct neighbors, and $f(x):  \mathbb{X}
\rightarrow \mathbb{R}$ is a fitness function that gives to every
$x \in \mathbb{X}$ a proprietary quantity to be interpreted as a
'quality' information or fitness~\cite{richengel14,stad03}. In this definition, the configuration space together with the neighborhood structure expresses a (multi--dimensional) 'location', while fitness is a property of this location. 
The configuration space itself can be seen as  an unordered (finite or infinite) list of configurations that the genetic specifications of biological systems can have. The neighborhood structure defines a  (possibly multi--dimensional) order of this list by establishing what is directly next to each element of the configuration space. As direct neighbors of an element have a neighborhood structure themselves, this naturally establishes distant neighbors of the element as well. 

The definition of a (static) landscape has the consequence of each configuration possessing a constant fitness value.  For several reasons this might not realistically  reflect the evolutionary scenario to be described and generally may restrict the descriptive power and versatility of the landscape model. Hence,
assuming that fitness may change over time, while configuration space and neighborhood structure remain constant, the  definition above can be extended to a dynamic fitness landscape, which can be expressed as the quintuple $\Lambda_D=(\mathbb{X},n,\mathcal{K},F,\phi)$~\cite{rich14a}. In addition to the elements of the static landscape, there is the time set $\mathcal{K}$, the set of all fitness function $F$ in time $\kappa \in \mathcal{K}$, and the transition map $\phi$ defining how the fitness function changes over time.  It is noteworthy that for a discrete time set $\mathcal{K}$, for instance for the integer numbers $\mathcal{K}= \{0,1,2,\ldots \}$, the notion of a dynamic landscape coincides with the notion of a series of static landscapes. In other words, two static landscapes $\Lambda_S^{(1)}=(\mathbb{X},n,f^{(1)})$ and $\Lambda_S^{(2)}=(\mathbb{X},n,f^{(2)})$ can be reformulated as one dynamic landscape  with $(f^{(1)}, f^{(2)}) \in F$ and $\phi$ describing how $f^{(1)}$ changes into $f^{(2)}$.

\subsection{Player landscapes}
Applying a landscape approach for describing evolutionary dynamics requires to address what may constitute a configuration $x \in \mathbb{X}$ and its neighborhood $n(x)$, but also what defines fitness $f(x)$. 
For the evolutionary game described in the previous sections, there are several modelling options, which will be reviewed in the following. The actual modelling choice of $\mathbb{X}$, $n$ and $f$ and their interdependencies may additional entail a  landscape that is dynamic.
The simplest modelling choice is to equate configurations with players $\mathcal{I}$, which for $N$ players leads to a (finite) configuration space with $N$ elements.  The neighborhood structure follows from the $d$ coplayers that each player has, which can be $1 \leq d \leq N-1$.
 In other words, the neighborhood of a player consists of all the other players with which it is mutually engaged in a game according to the interaction matrix $A_I$. Hence, such a player landscape  $\Lambda_\mathcal{I}$ can be specified by $\Lambda_\mathcal{I}=(\mathcal{I},A_I,f)$. 
 A popular form of such player landscapes is to place the players on a two--dimensional square lattice and define the coplayers to be Von Neumann (or Moore) neighborhoods, which consists of the lattices cells orthogonally (or additionally diagonally--adjacent) surrounding a central cell. Such an arrangement fixes the number of direct neighbors to $d=4$ (or $d=8)$, but yields a convenient way of visualizing the quality information over the resulting two--dimensional structure, which might be one reason for the popularity of these neighborhoods. 
  The most obvious choice of the quality information is payoff $p$ or quantities directly derived from it such as the linear fitness $f=1+\delta \cdot p$ introduced earlier. 
This has led to label such landscapes as payoff landscapes~\cite{brede11}.

 There are, however, several problems with such a player landscape model. The main problem is that with the player's and coplayers' strategies two quantities decisive for the amount of payoff are not directly attached to the configuration, which is player only. Strategies can be seen as ambiguous and polyvalent properties of the configuration of players. This means that the payoff attributable to a configuration depends on both the  player's strategy and also on the  strategies of its neighboring coplayers. This aspect is known as frequency--dependence as the payoff can be seen as to depend on how frequent the strategy that the player adopts also occurs in the coplayers. In short, fitness derived from payoff is dynamic so that the player landscape $\Lambda_\mathcal{I}=(\mathcal{I},A_I,\mathcal{K},f(\kappa),\phi)$ is dynamic as well. Moreover, the dynamics of $f(\kappa)$ is caused not only by frequency--dependence, but also by strategy updating for which the player landscape model does not directly account and  both these causes can hardly be separated from each other. Hence, the transition map $\phi$ describing how the fitness $f(\kappa+1)$ relates to $f(\kappa)$ is not straightforwardly definable. 
In addition, modelling configurations of a landscape by players means that the neighborhood structure is the adjacency matrix. In other words, a variable network of interaction, as in coevolutionary games, implies a changing neighborhood structure. To conclude a player landscape of a coevolutionary game involves changing neighborhood structure as well as  dynamic fitness. This may make analyzing such a landscape rather complicated.
 
There is another reason for the difficulties to deduce meaningful conclusions from 
 payoff over a  player landscape. Topological features of the landscape can be used as a starting point for estimating how likely transitions from low--fitness configurations to high--fitness configuration are and also which configurations are most likely to be a  steady state of evolutionary dynamics.   However, which player in a symmetric game as specified by the payoff matrix (\ref{eq:payoff}) exactly is  a likely high--fitness  outcome of an evolutionary process is  not very relevant. 
A much more important question is what fraction of the players in the long run settles stably to one of the possible strategies. In pursuing this question, there are several works that define the quality information of the landscape to be the strategy to which a player temporary or finally settles. This means the 'fitness' is expressed by the strategy vector $\pi(k)$. The results have been visualized by coloring the players according to their strategy~\cite{nowak93,nowak04}. Such a modelling has the advantage that the spatial and temporal distribution of the strategy preference can be visualized with respect to the player--coplayer structure. However, payoff--based fitness as the main drive of evolutionary game dynamics is not an explicit component of such a landscape. 

\subsection{Strategy landscapes}

An alternative modelling choice
is to equate configurations with all possible combinations of strategies that each player and its coplayers can have. An element $\pi \in \Pi$ of the strategy configuration space $\Pi$ is comprised of the strategies of any player $\mathcal{I}_i$, $i=1,2,\ldots,N$, and its up to $N-1$ coplayers: $\pi=(\pi_1\pi_2 \ldots \pi_N)$. Note that the strategy configuration space  $\Pi$ generalizes the time--dependent strategy vector $\pi(k)$ towards all of its possible instances. 
  Hence, for $N$ players with two possible strategies,  $\Pi$ contains $\ell=2^N$ elements. If we binary code the strategies cooperation and defection  (for instance $C_i=1$, $D_i=0$), an element  $\pi \in \Pi$ appears as binary string of length $N$. Note that the bit--count of $\pi$, $\text{bc}(\pi)$, provides a simple way of expressing the fraction of cooperators $\frac{\text{bc}(\pi)}{N}$. 
It is assumed that only one player or coplayer  can change its strategy at a given point in time. This implies
a
 neighborhood structure where  each element $\pi$ has  $N$ direct neighbors which are   distanced by 
Hamming distance  of $1$, which is denoted by $\mathcal{H}_d^1$. With such a modelling we obtain for payoff--based fitness $f$ 
a unique and static landscape $\Lambda_{\Pi}^i=(\Pi,\mathcal{H}_d^1,f_i)$ for each player $\mathcal{I}_i$ and each network of interaction. As the game specified by the payoff matrix (\ref{eq:payoff})
is symmetric,  the strategy landscapes $\Lambda_{\Pi}^i$ are topologically alike for all players $\mathcal{I}_i$. They can be transformed into each other by shifting and reshuffling.  For a landscape interpretation this topological likeness implies that landscape quantifiers such as the number of maxima, or correlation structure, or information content do not vary over the $\Lambda_\Pi^i$.

\begin{figure*}[t]

\includegraphics[trim = 38mm 65mm 40mm 100mm,clip, width=4.3cm, height=4cm]{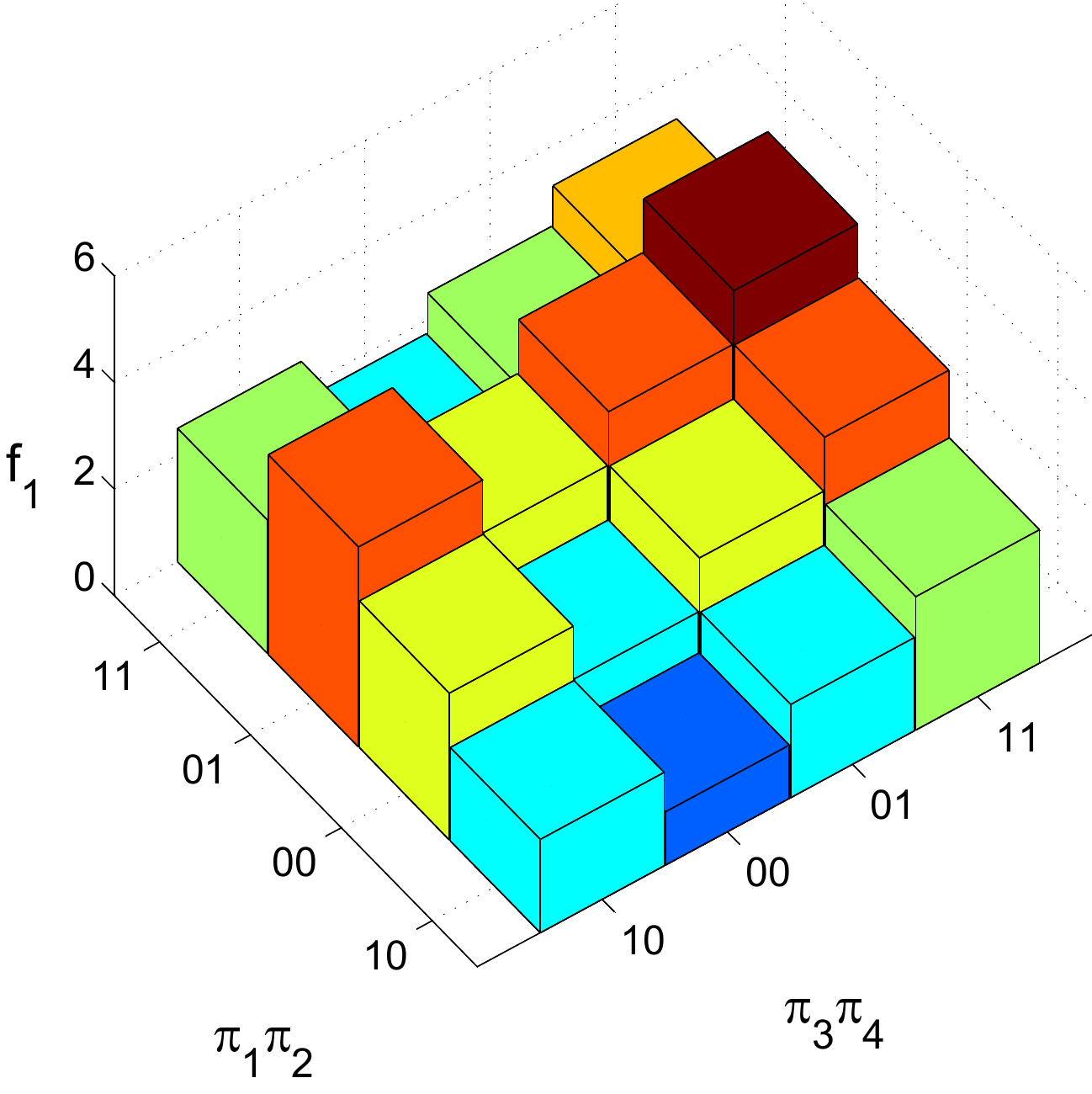} 
\includegraphics[trim = 38mm 65mm 40mm 100mm,clip, width=4.3cm, height=4cm]{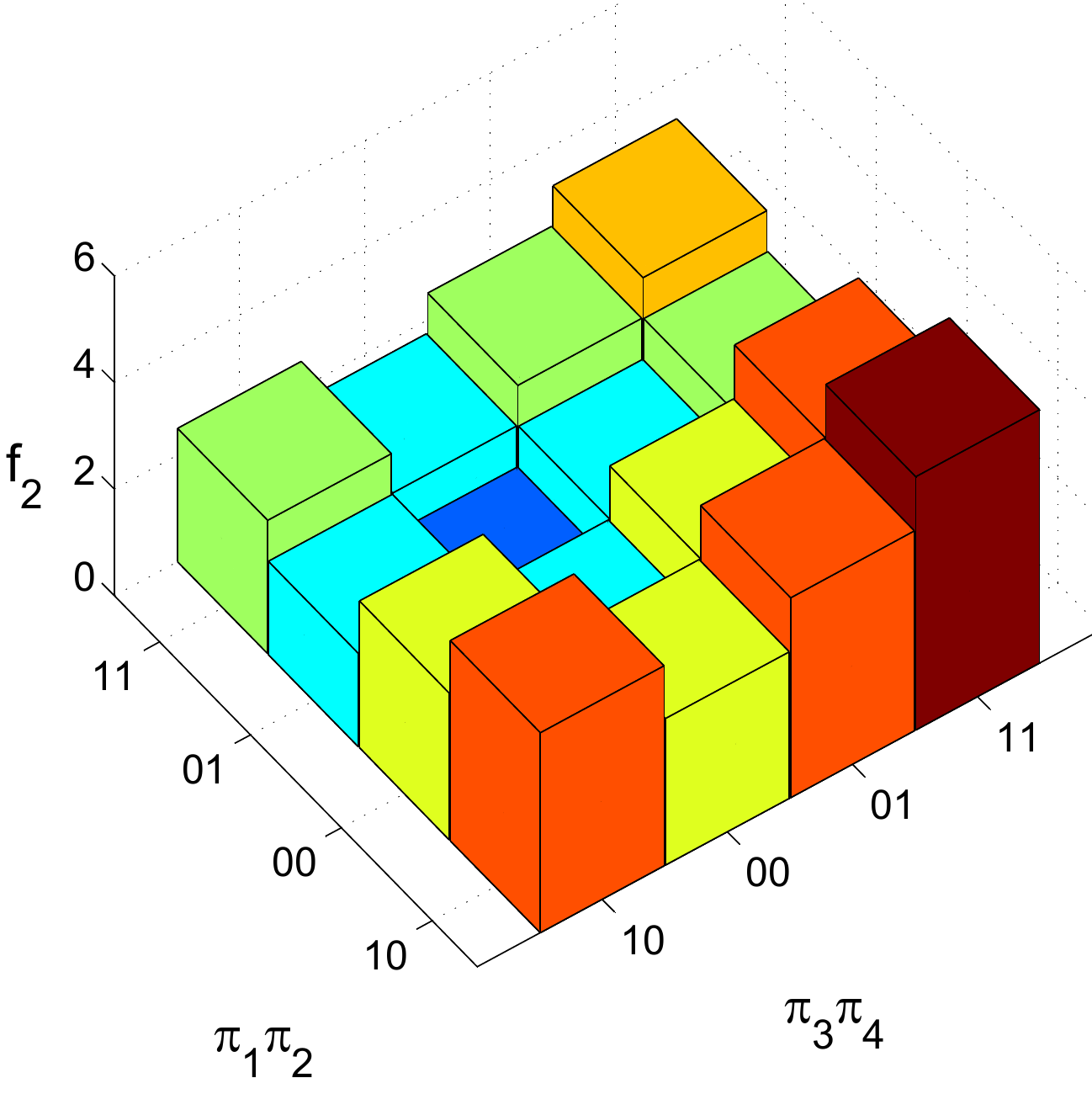} 
\includegraphics[trim = 38mm 65mm 40mm 100mm,clip, width=4.3cm, height=4cm]{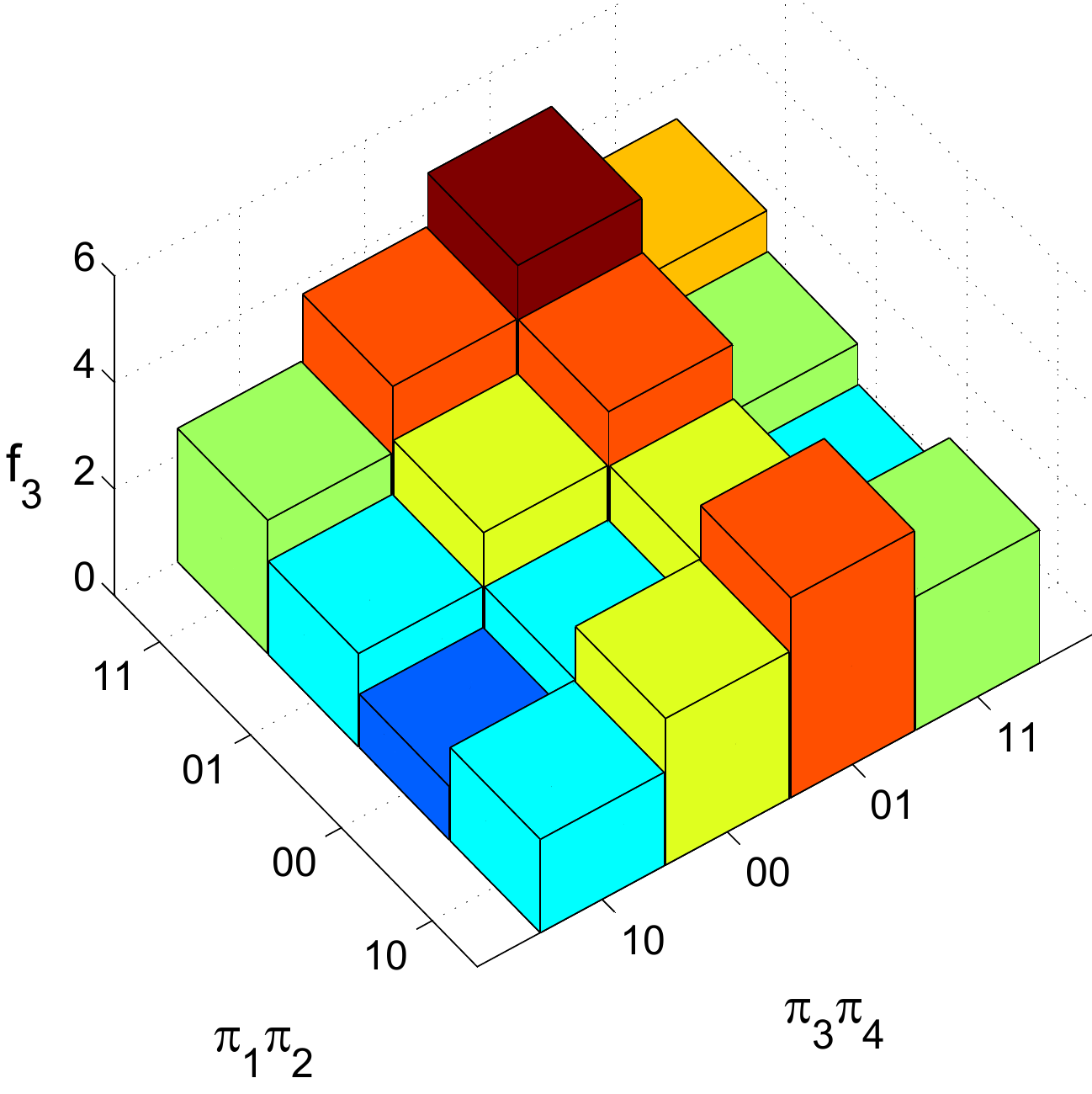} 
\includegraphics[trim = 38mm 65mm 40mm 100mm,clip, width=4.3cm, height=4cm]{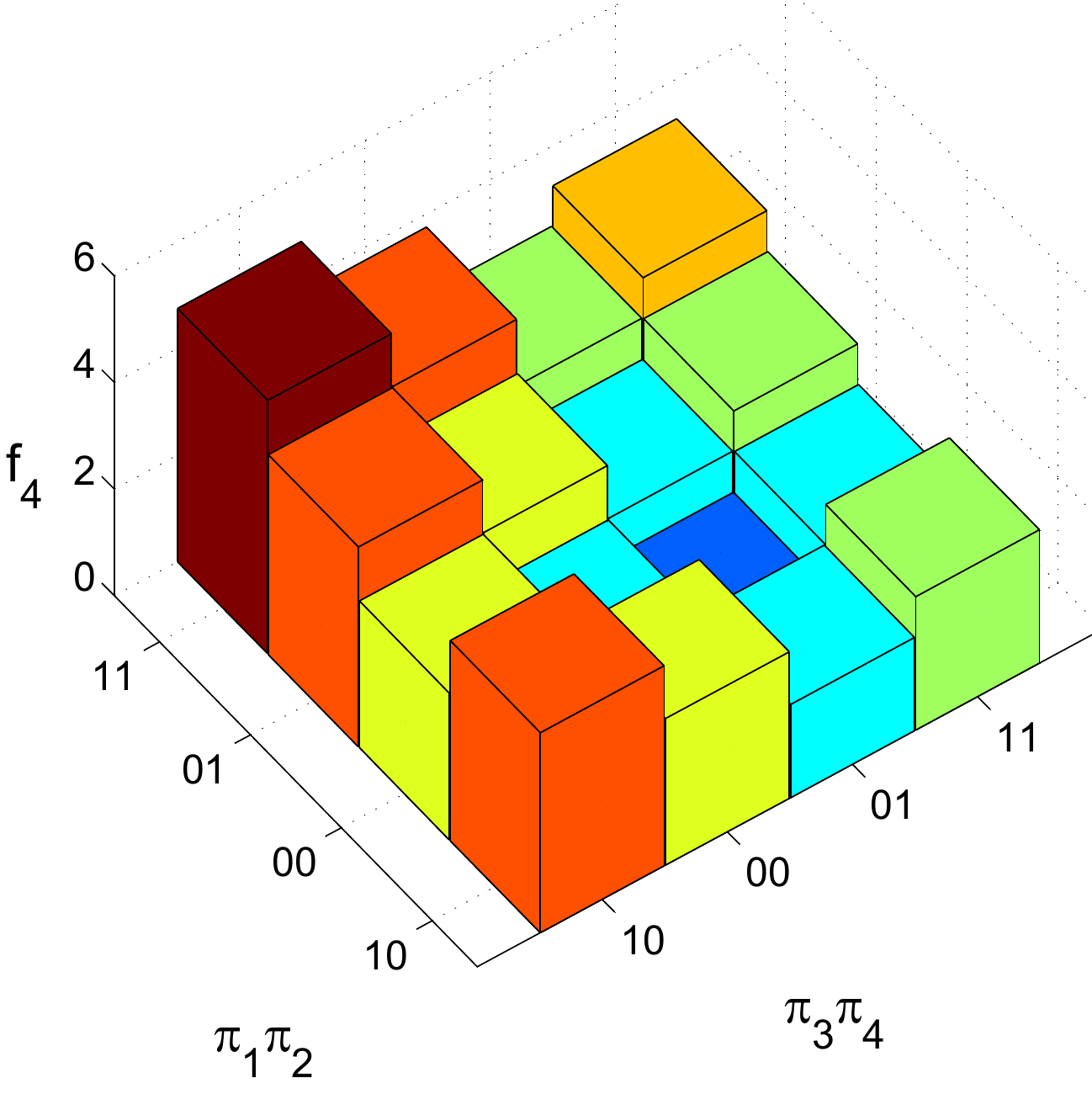}

\caption{Strategy landscapes $\Lambda_{\Pi}^i$ for a  PD game with $N=4$ and a complete network of interaction.   Same colors give equal fitness values $f=1+\delta p$ for payoff $p$ with $\delta=0.25$. Each strategy configuration $\pi=(\pi_1 \pi_2 \pi_3 \pi_4)$ has $N=4$ neighbors distanced by Hamming distance $\mathcal{H}_d^1$, while periodic boundary conditions apply.  }
\label{fig:n4}
\end{figure*}

For $N=4$, the landscapes   can be visualized in two dimensions, see Fig. \ref{fig:n4}. It shows  $\Lambda_{\Pi}^i$,  $i=1,2,3,4$, for  the payoff matrix
$\bordermatrix{~ & C_j & D_j \cr
                  C_i & 3 & 0 \cr
                  D_i & 5 & 1 \cr}$  and   $A_I=\begin{pmatrix} 0 & 1 & 1 & 1 \\ 1 &  0 & 1 & 1 \\ 1 & 1 & 0 & 1\\ 1 & 1 & 1 & 0 \end{pmatrix}$ the adjacency matrix specifying a PD game with  a complete  network of interaction and $d=3$. 
We find that $\mathcal{L}_3(4)=1$ and hence the game is static with respect to updating the network of interaction. Observe that for each player  there is only one maximum fitness value (the player is defecting, while all coplayers cooperate) and one minimum fitness value (the player cooperates, while all coplayers defect). Apart from the single maximum and the single minimum, there are several configurations that have the same fitness value. Interestingly, these configurations do not form a neutral network~\cite{richengel14} as they are no direct neighbors.   From the strategy landscape $\Lambda_{\Pi}^i$ it can be concluded which strategy for the player $\mathcal{I}_i$  (with respect to the strategies of the coplayers) yields the highest fitness and is therefore most desirable from the perspective of $\mathcal{I}_i$. Nonetheless, the evolutionary path from a given start configuration may depend on, and be influenced by, the strategies provided to and/or received from the coplayers. In addition, from the perspective of another player, another strategy configuration is best. Best configurations for respective players, however, are mutually exclusive, which is a defining feature of social dilemma games such as PD.   Consequently, each strategy landscape $\Lambda_{\Pi}^i$ can be seen as a building block that constructs a strategy landscape of the game.
Such a game landscape would allow conclusions as to what strategy configurations are preferred if  all players and their interactions are taken into account. In other words, a game landscape   may model the dynamics caused by the strategy updating processes discussed in Sec.    \ref{sec:statupdate}.

\subsection{Game landscapes}

Reconsider the game with $N=4$ players, for which Fig. \ref{fig:n4} depicts the player--wise strategy landscapes  $\Lambda_{\Pi}^i$. At every point in time $k$, the game can be seen as occupying one of its $2^4=16$ configurations. Put another way, the actual strategy vector $\pi(k)$ specifies an actual configuration on the landscapes $\Lambda_{\Pi}^i$. For each player $\mathcal{I}_i$, its landscape     $\Lambda_{\Pi}^i$ gives its actual fitness $f_i(k)$.  The strategy updating process means that one player provides its strategy for another player to receive. The receiving player changes its strategy. According to the landscape view this process corresponds with changing the actual configuration $\pi(k)$ to a neighboring configuration $\pi(k+1)$. The change of configuration entails that all players may experience a change of fitness as well. No player can hold onto its configuration if the strategy updating process is underway unless one of the two absorbing configurations are reached, namely $\pi=(00\ldots00)$ or $\pi=(11\ldots11)$. 
In the following, the strategy updating processes birth--death (BD) and death--birth (DB) will be discussed. For these processes transition probabilities can be derived~\cite{patt15}. Furthermore,  it will be convenient to rewrite the landscape $\Lambda_{\Pi}^i$ as its decomposition $\Lambda_{\Pi}^i= \{ \lambda_\ell^i\}$, $\ell=1,2,\ldots,2^N$, where each $\lambda_\ell^i$ contains the fitness and neighborhood of configuration $\ell$. 

Assume that the game is in the configuration $\pi(k)=(1101)$, which means that player $\mathcal{I}_3$ is defecting, while the three other players are cooperating. According to the PD game,  the fitness of $\mathcal{I}_3$  is highest, the three other players have the same (albeit lower) fitness. 
To start with, let us consider a BD strategy updating. A player's strategy is chosen at random with a probability proportional to fitness to be a source (hence birth). The birth probability of a configuration $\ell$ of player $\mathcal{I}_i$ therefore scales to $b_\ell^i= \frac{\lambda_\ell^i}{ \sum_{\ell=1}^{2^N} \lambda_\ell^i}$, where the  $\lambda_\ell^i$ are the decompositions of the landscape $\Lambda_{\Pi}^i$.  The player with the highest fitness is most likely to be a source, 
 which is $\mathcal{I}_3$ with strategy $\pi_3(k)=0$. 
Which one of the three players is the target  to receives the strategy (hence death) is due to chance but influenced by possible restrictions regarding the replacement. Hence, the death probability of a player  $\mathcal{I}_i$  is $d_i=\frac{1}{N}\displaystyle \sum_{j=1}^{N} \frac{w_{ji}}{\sum_{i=1}^{N} w_{ji} }$, where the  $w_{ji}$ are the elements of the  replacement matrix $W_R$ possibly restricting replacement of strategies as discussed in Sec.  \ref{sec:statupdate}. Note that the death probability is independent of fitness and hence the same for all configurations.  A  BD (and also a DB) updating does not envisage self--replacement and hence the replacement matrix $W_R$ must have diagonal elements $w_{ii}=0$. Hence, if there are no replacement restrictions, then $d_i=\frac{1}{N}$ for all players using a BD updating.
 Assume that all players can be a target and  $\mathcal{I}_2$ is chosen. Hence, the strategy configuration after the  strategy updating is $\pi(k+1)=(1001)$. The players $\mathcal{I}_2$ and $\mathcal{I}_3$ have leveled their fitnesses, while the fitness of both the other players is fallen even more. Now consider a DB strategy updating. Here, a player's strategy is chosen at random with a probability proportional to the inverse of its fitness to be a target (hence death).  Therefore, the death probability of  a configuration $\ell$ of player $\mathcal{I}_i$ can be expressed as scaling to $d_\ell^i= 1-\frac{\lambda_\ell^i}{ \sum_{\ell=1}^{2^N} \lambda_\ell^i}$. Still assume that the game is in the configuration $\pi(k)=(1101)$ and as the players $\mathcal{I}_1$, $\mathcal{I}_2$, and $\mathcal{I}_4$ have the same (low) fitness values, one of them is most likely to be the target. Suppose  $\mathcal{I}_1$ is chosen. Which one of the three players provides its strategy  as a source (hence birth), depends on chance and possible replacement restrictions. We get the birth probability  $b_i=\frac{1}{N}\displaystyle \sum_{j=1}^{N} \frac{w_{ij}}{\sum_{i=1}^{N} w_{ij} }$, which is the same as the death probability in BD, but the target and the source are switched in the elements of the replacement matrix. Note that only if the player $\mathcal{I}_3$ is chosen, a change in the configuration takes place, that is the strategy configuration after the  strategy updating is $\pi(k+1)=(0101)$. In other words, the outcome of both a DB and a BD updating may be the same, but the probabilities to reach it may be different. 

\begin{figure*}[t]

\includegraphics[trim = 12mm 65mm 10mm 100mm,clip, width=4.3cm, height=4cm]{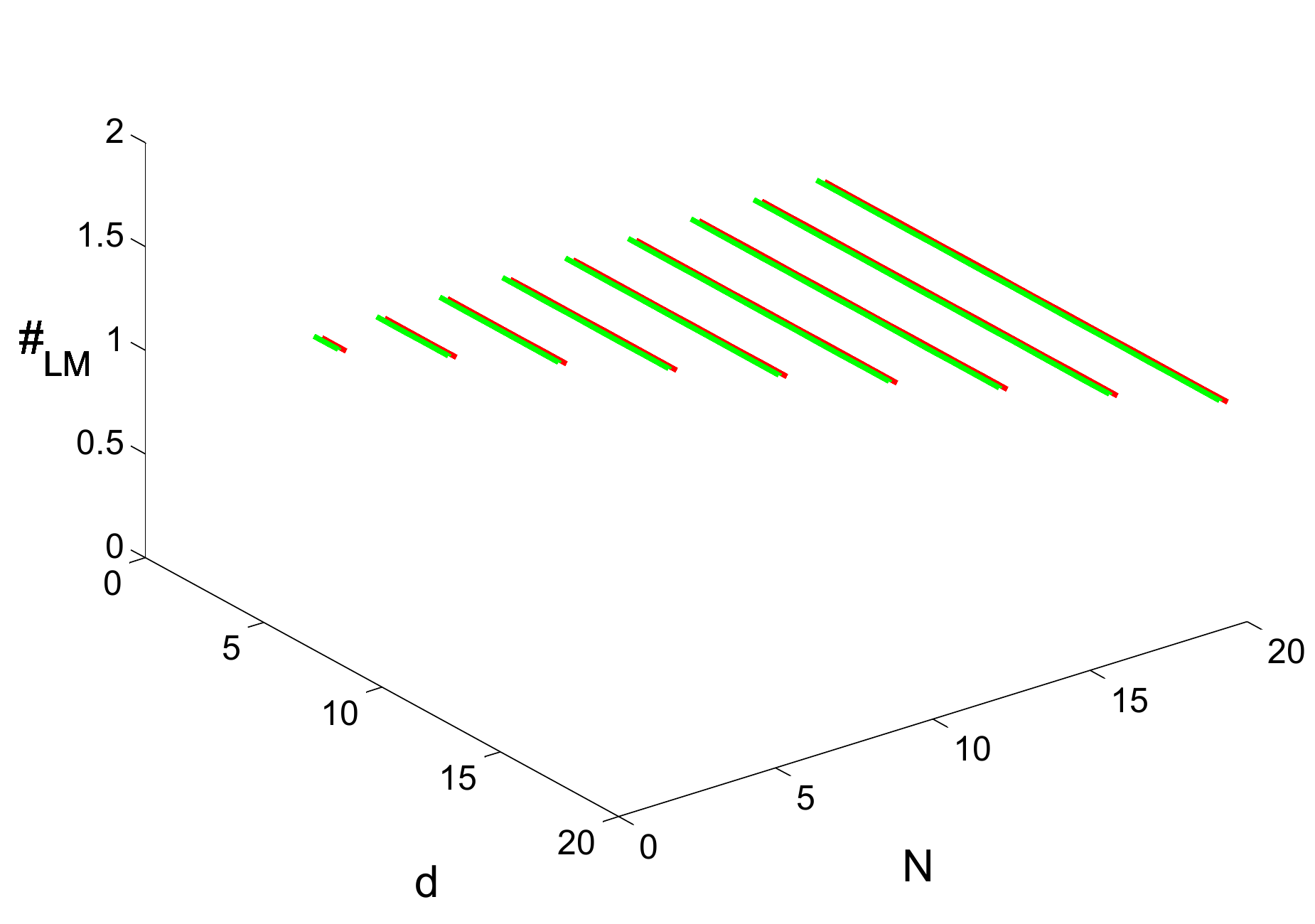} 
\includegraphics[trim = 12mm 65mm 10mm 100mm,clip, width=4.3cm, height=4cm]{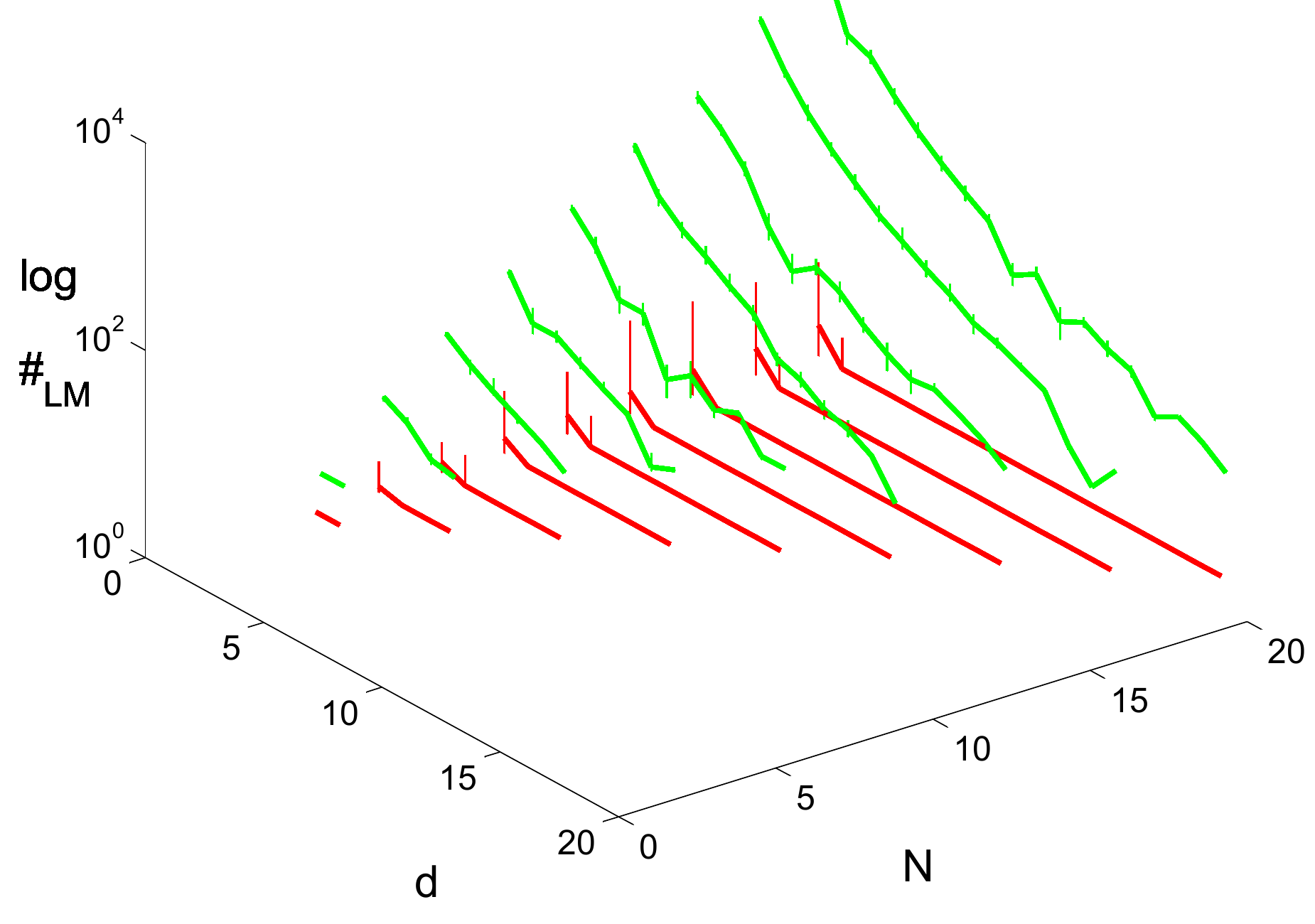} 
\includegraphics[trim = 12mm 65mm 10mm 100mm,clip, width=4.3cm, height=4cm]{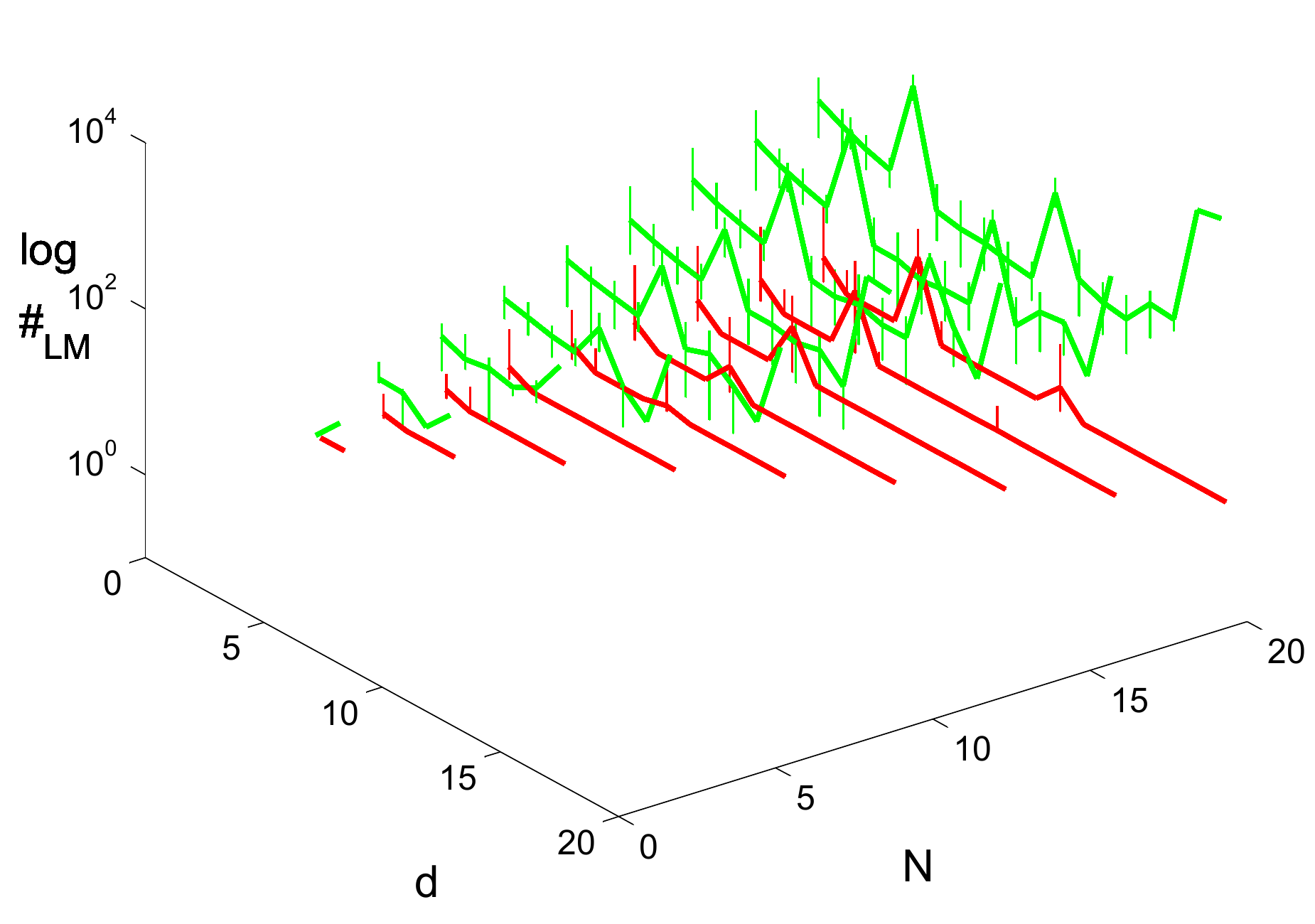} 
\includegraphics[trim = 12mm 65mm 10mm 100mm,clip, width=4.3cm, height=4cm]{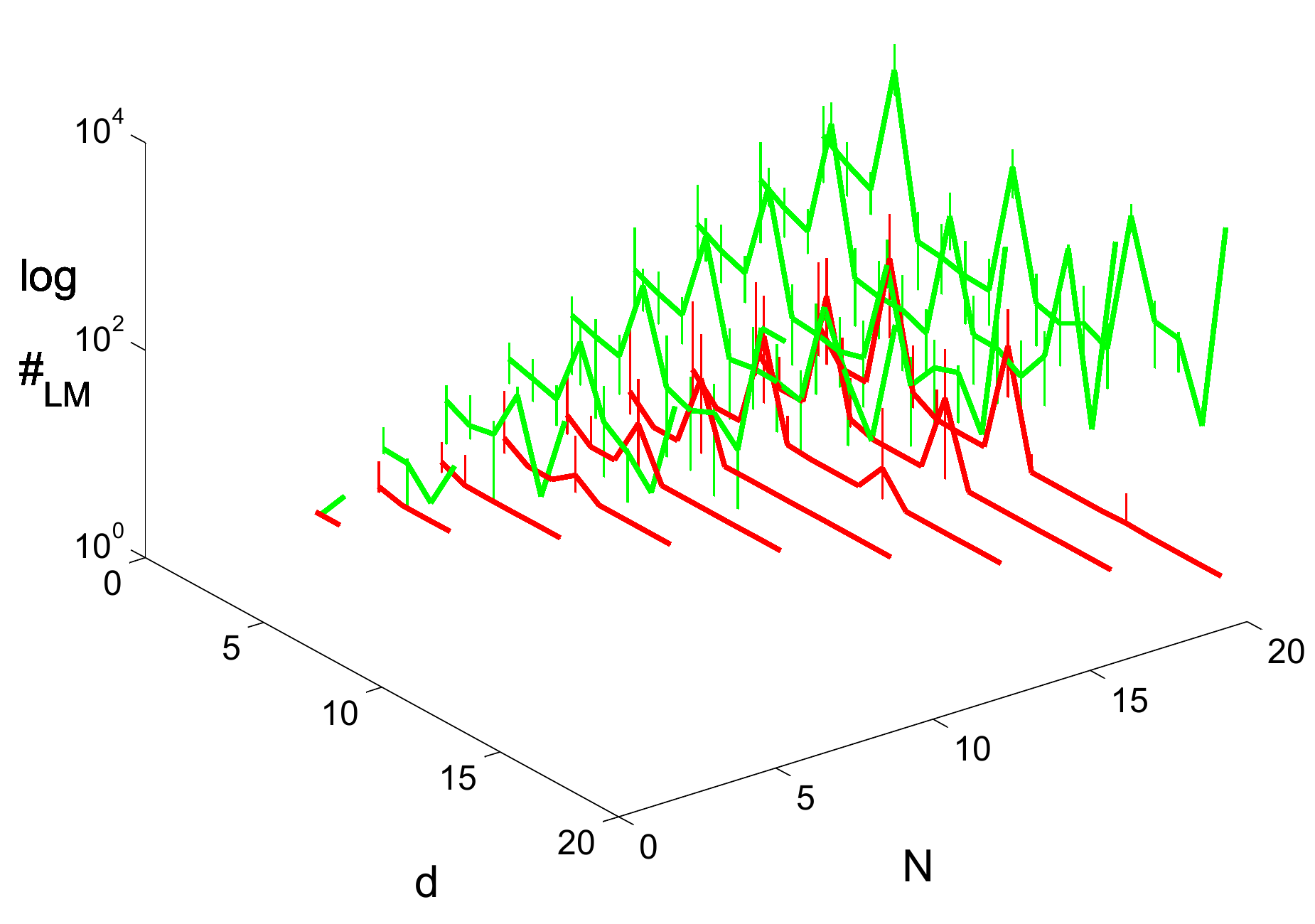} 

\hspace{1cm} (a) \hspace{4cm} (b) \hspace{4cm} (c) \hspace{4cm} (d) 

\includegraphics[trim = 12mm 65mm 10mm 100mm,clip, width=4.3cm, height=4cm]{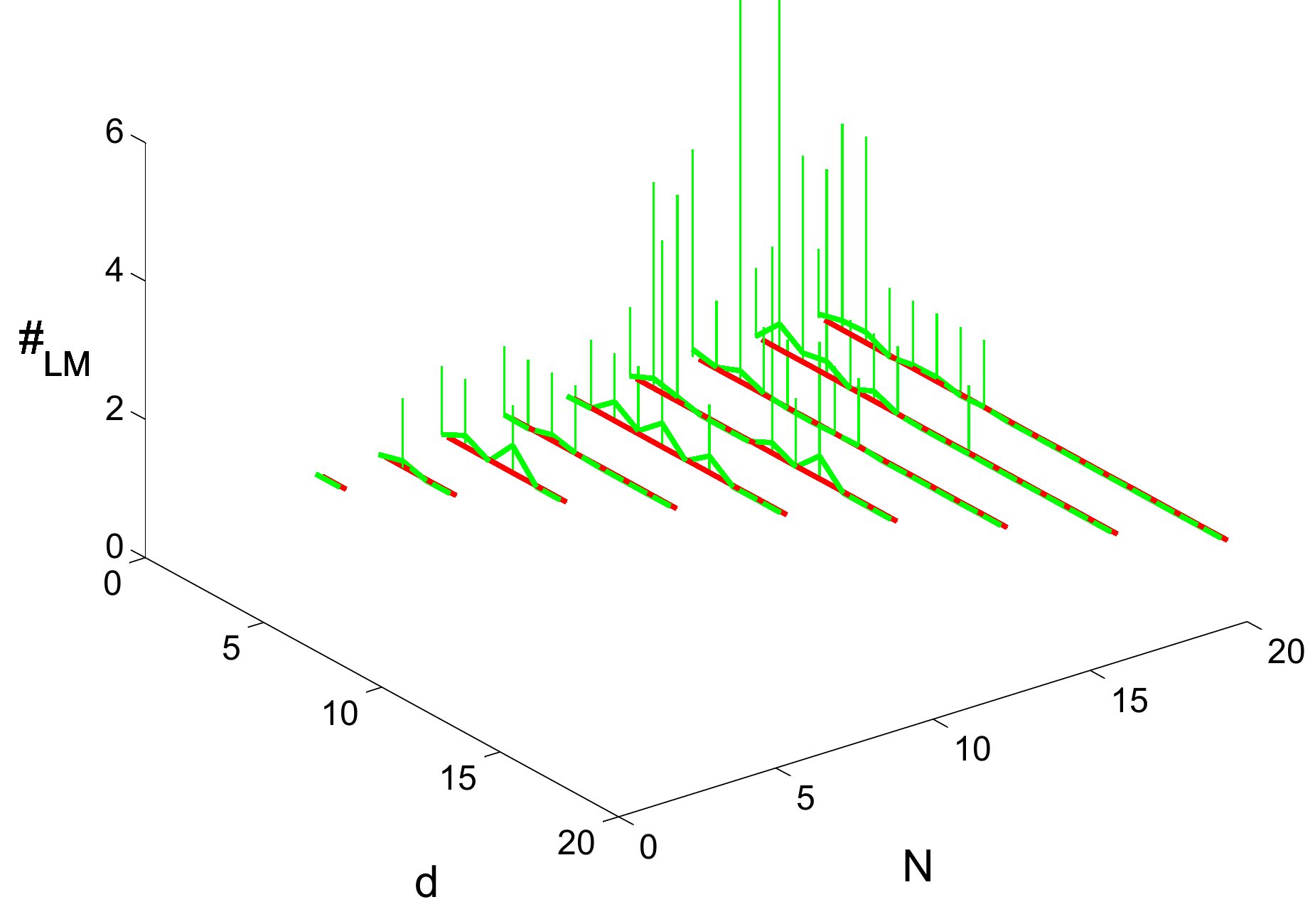} 
\includegraphics[trim = 12mm 65mm 10mm 100mm,clip, width=4.3cm, height=4cm]{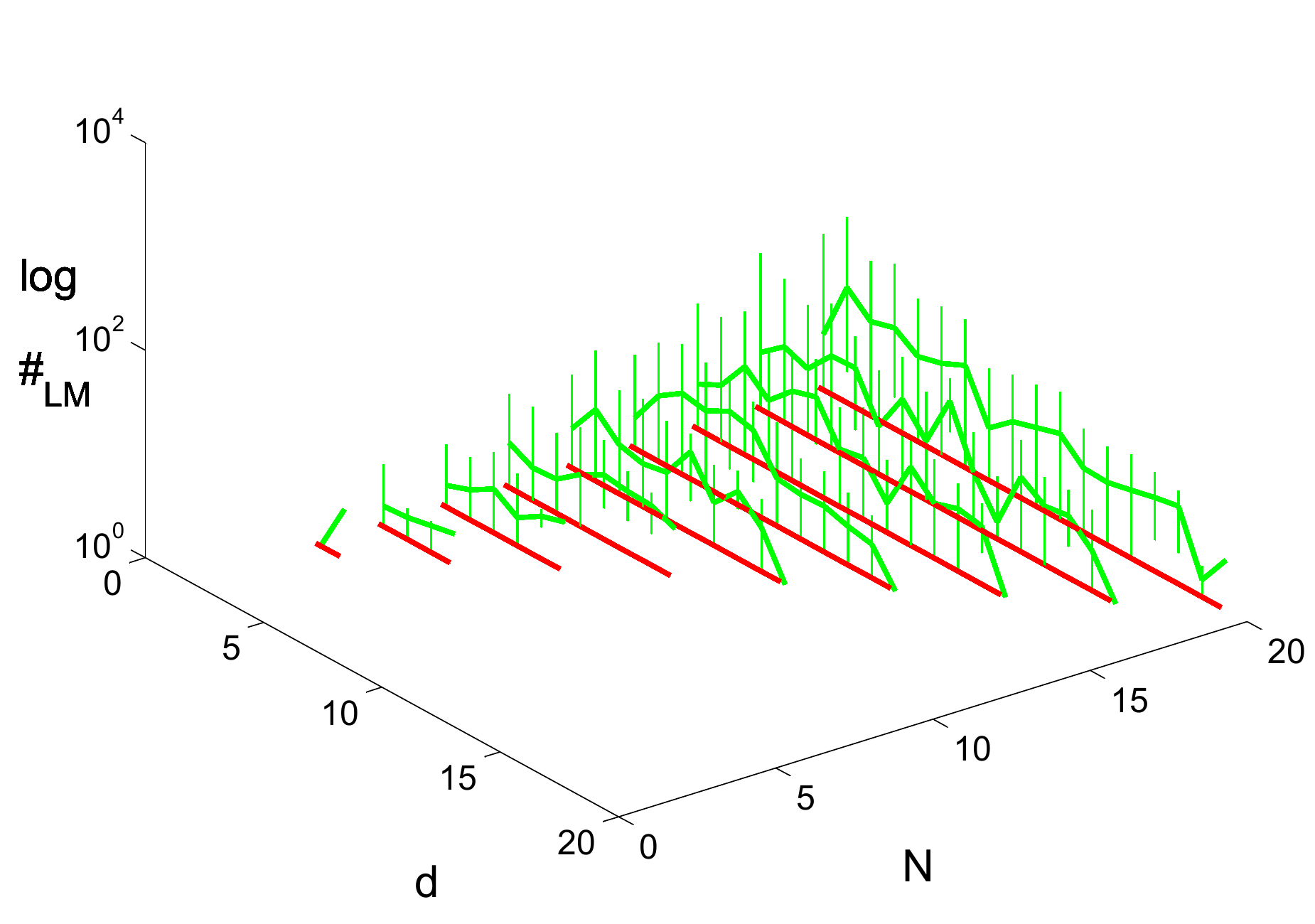} 
\includegraphics[trim = 12mm 65mm 10mm 100mm,clip, width=4.3cm, height=4cm]{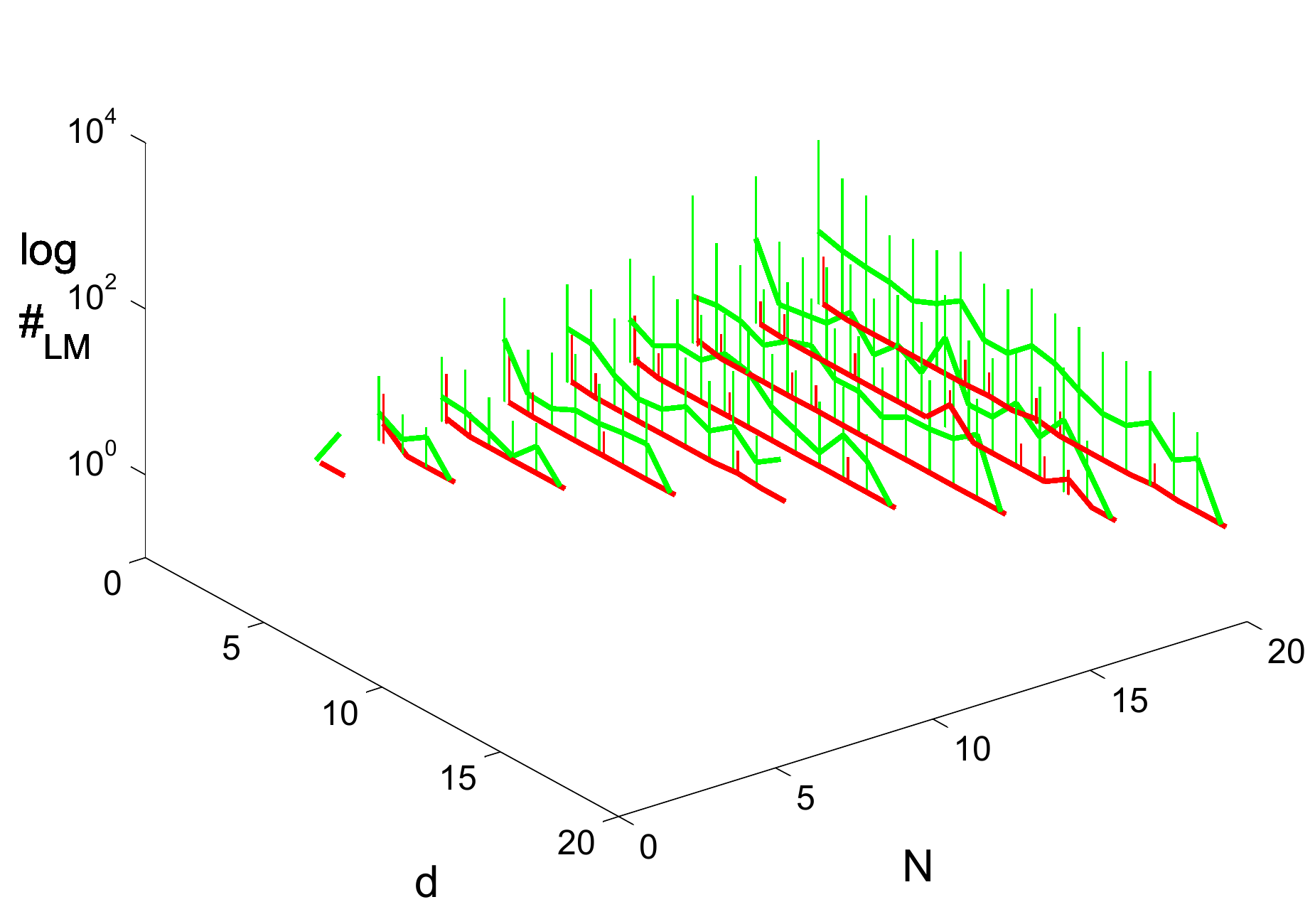} 
\includegraphics[trim = 12mm 65mm 10mm 100mm,clip, width=4.3cm, height=4cm]{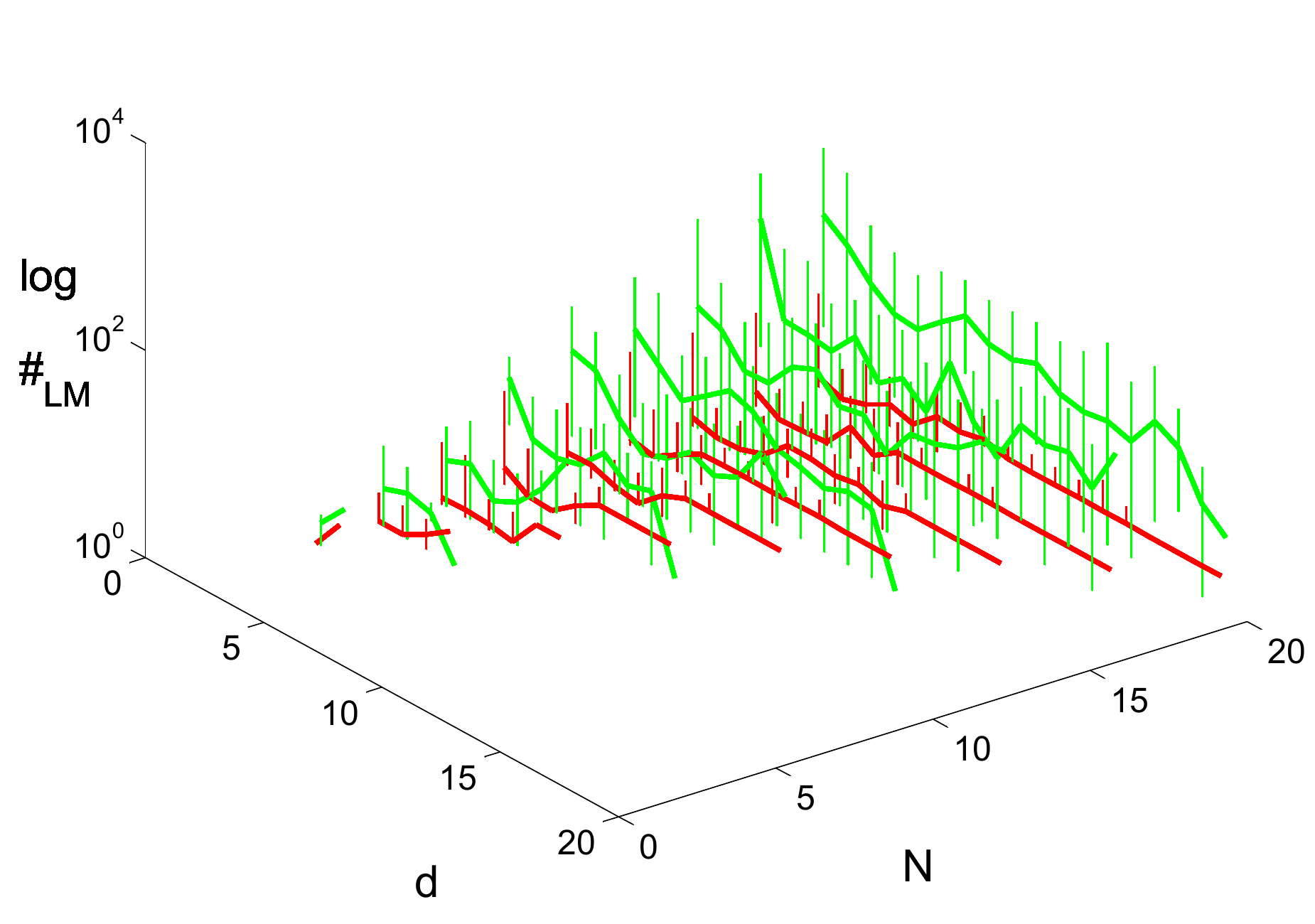} 

\hspace{1cm} (e) \hspace{4cm} (f) \hspace{4cm} (g) \hspace{4cm} (h) 
\caption{Modality measured by the number of local maxima $\#_{LM}$ over $N$ and $d$. Red gives the results for BD updating, green for DB updating. (a) -- (d): no replacement restriction.  $T=5$: (a) PD game,  (b) SD game; $T=7$: (c) PD game,  (d) SD game. (e) -- (h):  replacement restrictions.   $T=5$: (e) PD game,  (f) SD game; $T=7$: (g) PD game,  (h) SD game.   }
\label{fig:modality}
\end{figure*}

For a sufficient large number of updating processes (and therefore changes of configuration), there will be some configurations that the game occupies more often than others.   These may be the absorbing configurations with a bit count $\text{bc}(\pi)=0$ and $\text{bc}(\pi)=N$. Whether or not these absorbing configurations are reached and how long this takes, gives raise to fixation probabilities and times to fixation, which are interesting quantities to study~\cite{lieb05,patt15,sha12}. However, here it is focused on the question of how frequent any configuration is. 
The frequency of reaching a configuration
 scales to the probabilities of  birth and death discussed so far. Hence, for a BD updating the game landscape
\begin{equation}\Lambda_{\Pi}^{BD} =  \{ \lambda_\ell^{BD}\}=\left \{  \frac{1} { 1+\exp{\left(  \frac{1}{N} \sum_{i=1}^{N}    b_\ell^i d_i  \right) } }\right \}, \label{eq:bd} \end{equation}
can be defined, while for a DB updating, we set
\begin{equation} \Lambda_{\Pi}^{DB} = \{ \lambda_\ell^{DB}\}=\left \{  \frac{1} { 1+\exp{\left(  \frac{1}{N} \sum_{i=1}^{N}    d_\ell^i b_i  \right) } }\right \}, \label{eq:db} \end{equation}
both with $\ell=1,2,\ldots,2^N$. Different updating processes cast different game landscapes $\Lambda_{\Pi}^{BD}$ and  $\Lambda_{\Pi}^{DB}$ out of the same strategy landscapes of the players $\Lambda_{\Pi}^i$ . Given that the $\Lambda_{\Pi}^i$  are topological alike, and hence might be seen as to possessing symmetry properties, different strategy updating rules break the symmetry of the player--wise strategy landscapes.

The discussion so far has been for a constant network of interaction, that is a specific matrix $A_I$. As pointed out in Sec.  \ref{sec:netwupdate}, network updating can be understood as to be describable by a series of adjacency matrices $A_I(\kappa)$.
Hence, as the 'genetic description'  of the evolutionary game comprises of the strategy vector {\em and} the network of interaction,  the strategy configurations made up by the space $\Pi$ could be augmented by interaction configurations built by all possible networks of interaction. 
Consequently, the $\mathcal{L}_d(N)$ different interaction graphs enumerated by Eq. (\ref{eq:LDN}) could be seen as  configurations according to the landscape definitions discussed above. However, in view of the rather large number of possible graphs for a given $N$ and $1<d<N-2$ (a rough estimate of Eq. (\ref{eq:LDN}) for $d \ll N$ yields $\mathcal{L}_d(N) = \mathcal{O}( N^N)$), an alternative model is to understand different interaction graphs as dynamic instances of the strategy  landscape.  Put differently, the dynamics of the strategy landscape is the results of its variability with respect to the network of interaction. It should be noted that the timely order of the varying network of interaction could be interpreted as  temporal neighborhoods according to the neighborhood structure inherent to landscapes.  With network updating modelled as dynamic instances of player--wise strategy landscapes, we get a  dynamic landscape
$\Lambda_{\Pi}^i=(\Pi,\mathcal{H}_d^1,\mathcal{K},f_i(\kappa),\{ A_I(\kappa), A_I(\kappa+1)\})$ for describing a coevolutionary game. Apart from the strategy configuration $\Pi$ with neighborhood $ \mathcal{H}_d^1$ and the integer time set  $\mathcal{K}$, the quantity $f_i(\kappa)$ gives payoff--based fitness for each configuration, each player $\mathcal{I}_i$ and the $\kappa$--th network of interaction. The matrix pair $\{ A_I(\kappa), A_I(\kappa+1)\}$ of subsequent adjacency matrices specifies how the fitness $f_i(\kappa+1)$ relates to $f_i(\kappa)$. For these dynamic player--wise strategy landscapes, game landscapes for BD and DB updating can be defined according to the probabilities of birth/death and expressed as dynamic counterparts of Eqn. (\ref{eq:bd}) and (\ref{eq:db}). These dynamic BD and DB landscapes are the main topic of the numerical experiments reported in the next section.  

\section{Numerical experiments} \label{sec:num}
The numerical experiments with the dynamic fitness landscapes of coevolutionary games presented in this research consider a PD game and a SD game with  payoff matrices
$\bordermatrix{~ & C_j & D_j \cr
                  C_i & 3 & 0 \cr
                  D_i & 5 & 1 \cr}$ and
$\bordermatrix{~ & C_j & D_j \cr
                  C_i & 3 & 1 \cr
                  D_i & 5 & 0 \cr}$, respectively. Additionally, the effect of varying the $T/R$ ratio (which encourages  or dampens the temptation to defect) is studied. Therefore, results for $T=5$ are contrasted with $T=7$. The dynamics of the landscape is addressed by examining the effect of varying networks of interaction. Algorithms are employed that numerically generate  adjacency matrices $A_I(\kappa)$ specifying random regular graphs with given degree~\cite{bay10,blitz11,kim06}. It is checked if the graphs are connected. If a graph fails the check, there are isolated vertices that may bias controlling the interaction network via the graph's degree and hence such graphs are discarded. For the experiments, a set of up to $3000$ graphs with prescribed $N$ and $d$ are generated. For $\mathcal{L}_d(N)<3000$, the complete set of possible networks of interaction is used. Also, different replacement structures are analyzed.  The experimental setup follows previous works~\cite{ohts07} and defines the replacement matrix $W_R$ as random regular graph with given degree and guaranteed connectivity.  Additionally, the elements $w_{ij} \neq 0$ are filled with realizations of a random variable uniformly distributed on the interval $[0,1]$. The degree of $W_R$ is set to match the degree of $A_I$. All these experiments are carried out for BD and DB landscapes. Other updating schemes such as PC or IM can be treated likewise as for these processes transition probabilities are  also known~\cite{patt15} and hence  landscapes similarly to (\ref{eq:bd}) and (\ref{eq:db}) can be computed.  With the conventional PC--based  computational resources that were available, it was possible to experiment within a reasonable time--frame with up to $N=20$ players. All experiments employ a linear relationship $f=1+\delta p$ between payoff and fitness with $\delta=0.25$.

As pointed out before, the landscapes can only be visualized as two--dimensional projections up to $N=4$ players. For analyzing landscapes with more players, we need to resort to  landscape measures. As a first measure we look at modality expressed by the number of local maxima $\#_{LM}$.  Local maxima are potential steady states on the evolutionary path. Hence, the number of local maxima relates to the variety of possible evolutionary paths and consequently to the complexity of the evolutionary dynamics displayed. If there is just one (smoothly accessible) maximum, then all evolutionary paths converge to it and the evolutionary dynamics displayed is rather simple.  If, on the other hand, there is a large number of maxima, then the possible evolutionary paths may be very different resulting in more complex evolutionary dynamics. Fig. \ref{fig:modality} shows the  number of  local maxima $\#_{LM}$ over even $N$ and $d$. The red lines show a BD updating, the green lines a DB updating. In addition to the  $\#_{LM}$ averaged over the up to $3000$ different interaction networks tested (horizontal lines), the vertical spikes indicate the range between the least and the largest value of $\#_{LM}$. Except Fig. \ref{fig:modality}a and \ref{fig:modality}e showing the PD game for $T=5$, the  $\#_{LM}$ are given as logarithmic plots. 
The results show several  major trends that partly influence each other. A first is that SD  landscape have generally more maxima than PD landscapes and also DB updating generates more maxima that BD updating.
Further note that
$\frac{\#_{LM}}{2^N} \rightarrow 0$ for $N$ getting larger. 
All these results support previous findings about evolutionary games, for instance that PD games but also BD updating does not provide an advantage for cooperators~\cite{ohts07}. Thus, the small number of maxima of the player--wise landscapes $\Lambda_\pi^i$ (compare to Fig. \ref{fig:n4}) corresponds with the small number of maxima in the game landscape. By contrast, for the SD game and DB updating not only configurations where the defecting player earns the largest payoff are maxima of the game landscapes. Consequently, the number $\#_{LM}$ is larger. 
 Also, an increased $T/R$ ratio leads to more maxima, albeit lightly. In addition, for $T=7$ the spread of $\#_{LM}$ is over a larger range, indicating that different networks of interaction produces topologically different landscapes. The same applies for restrictions imposed by $W_R$, see Fig. \ref{fig:modality}e-h. Most visible for $T=5$, but also for $T=7$, replacement restrictions imply landscapes that vary substantially  over different networks of interaction.

\begin{figure*}[t]

\includegraphics[trim = 12mm 65mm 10mm 100mm,clip, width=4.3cm, height=4cm]{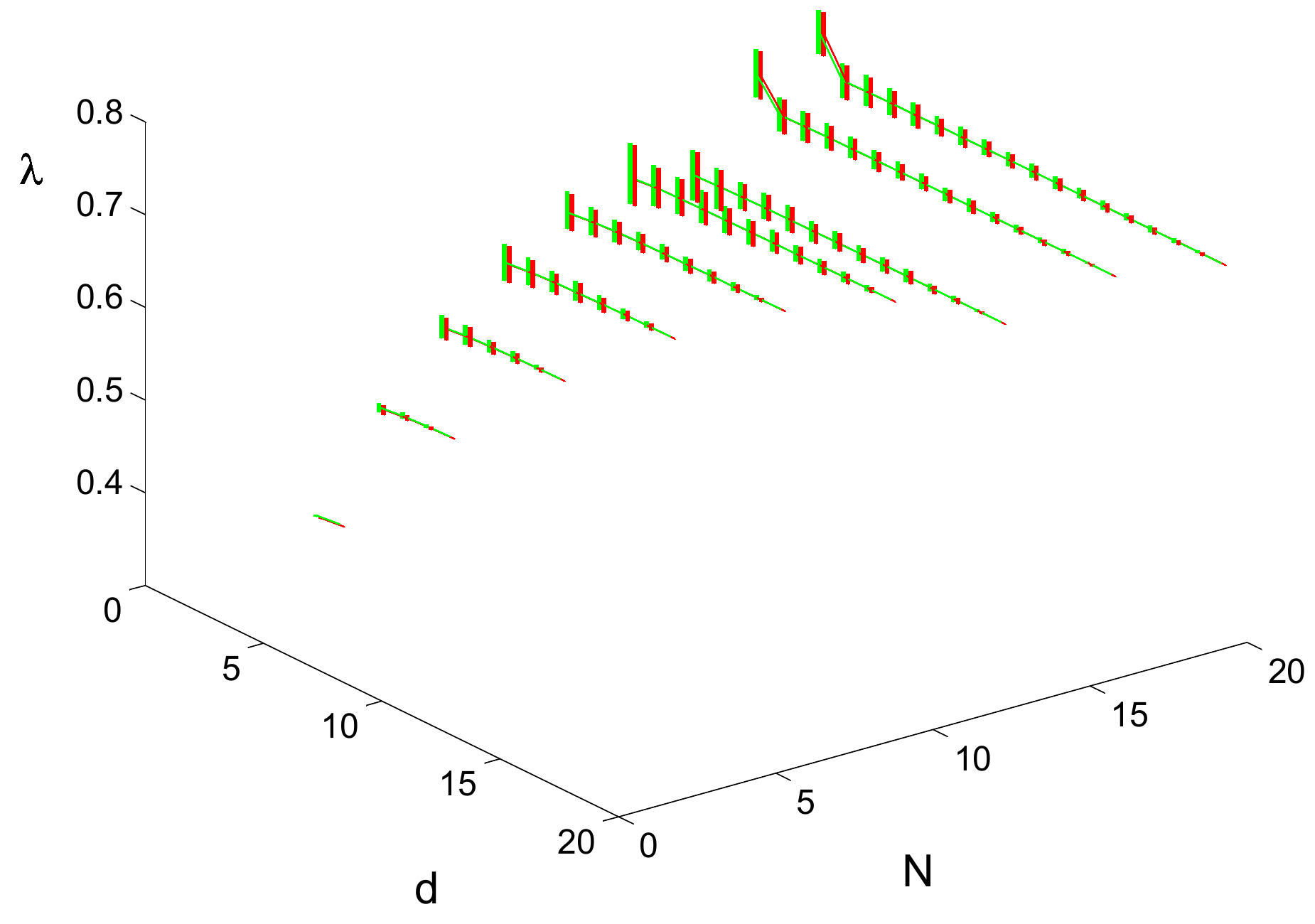} 
\includegraphics[trim = 12mm 65mm 10mm 100mm,clip, width=4.3cm, height=4cm]{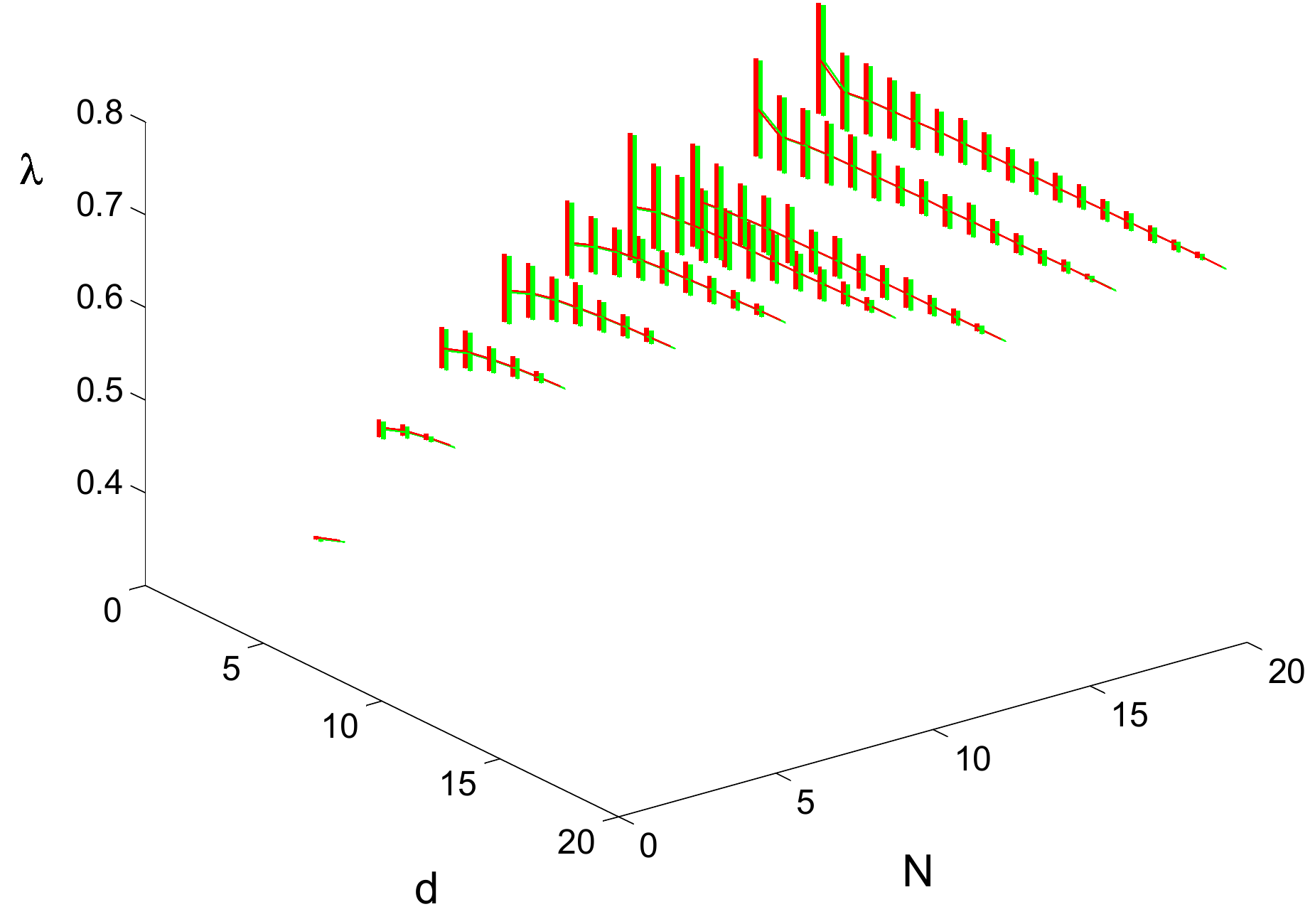} 
\includegraphics[trim = 12mm 65mm 10mm 100mm,clip, width=4.3cm, height=4cm]{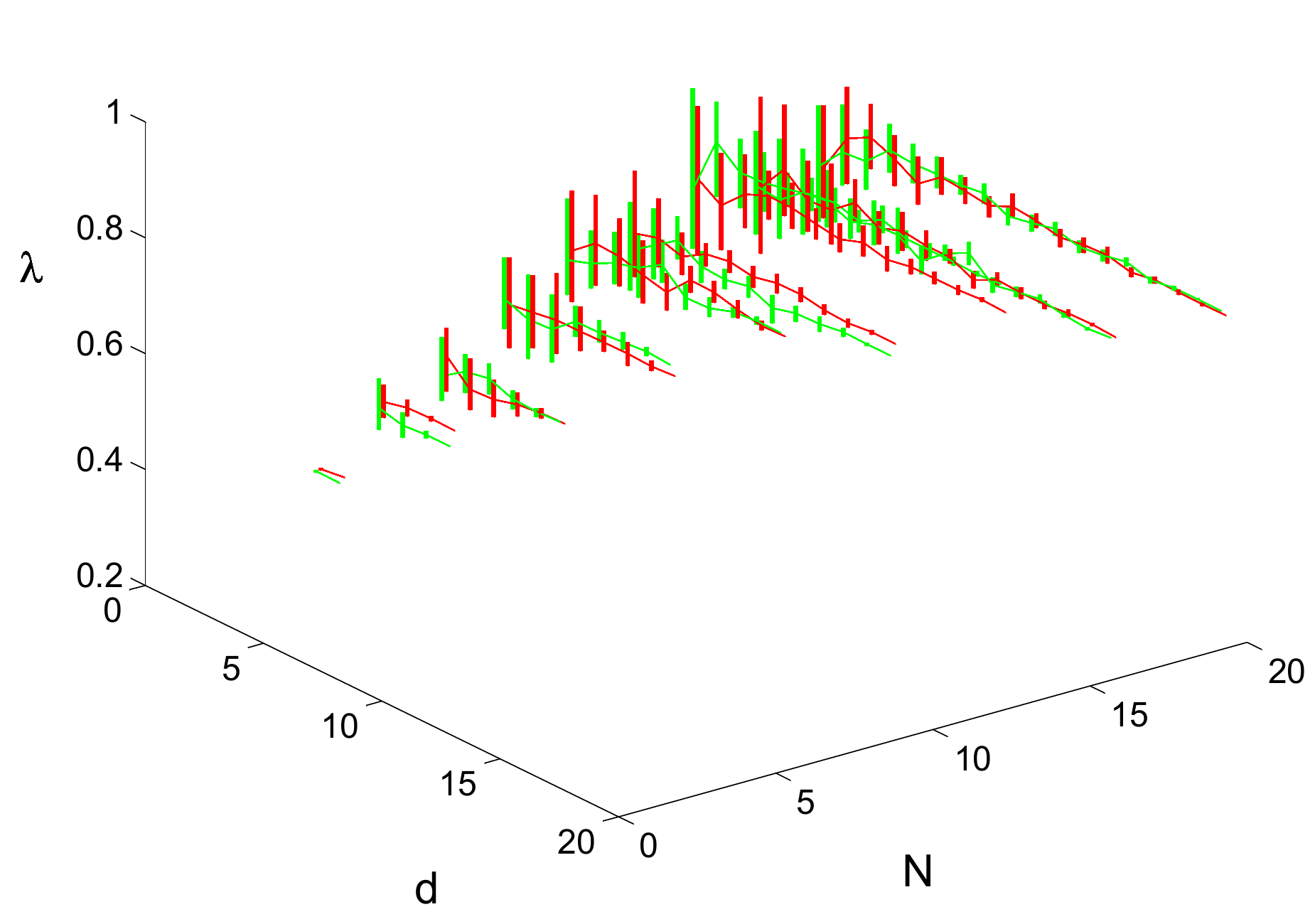} 
\includegraphics[trim = 12mm 65mm 10mm 100mm,clip, width=4.3cm, height=4cm]{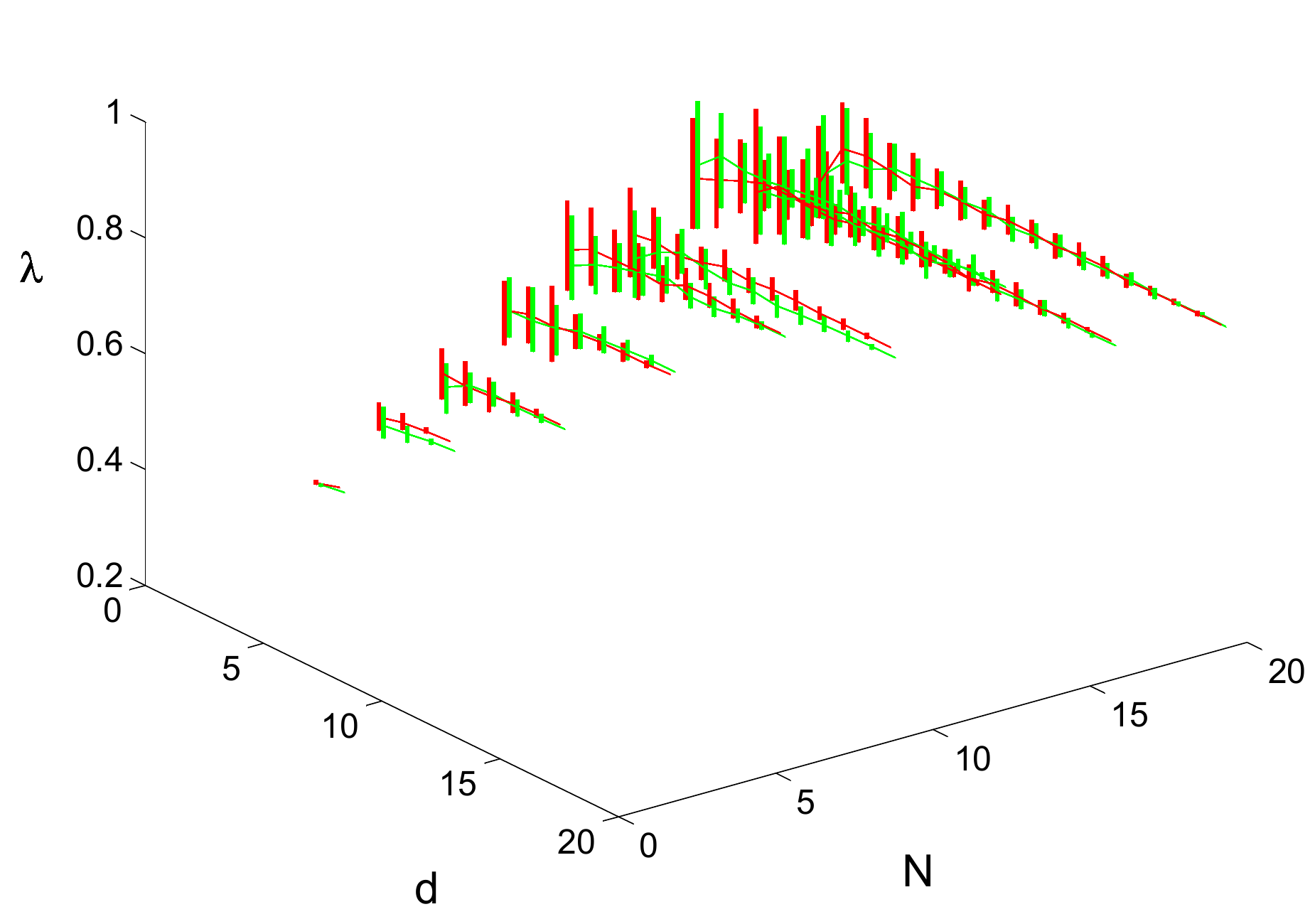} 

\hspace{1cm} (a) \hspace{4cm} (b) \hspace{4cm} (c) \hspace{4cm} (d)

\caption{Ruggedness measured by the correlation length $\lambda$ over $N$ and $d$.  Red gives the results for BD updating, green for DB updating. (a) -- (b): no replacement restriction.  (c) -- (d): replacement restriction. $T=5$: (a) (c) PD game,  (b) (d) SD game. }
\label{fig:rugg}
\end{figure*}

As a second landscape measure, the correlation length $\lambda$ is studied, see Fig. \ref{fig:rugg}. The correlation length $\lambda=-1/\log(|r(1)|)$ derives from the autocorrelation  $r(1)$ of time lag $1$ of the landscape's random walk correlation function~\cite{richengel14}. A random walk of length $10000$ was used, the results are averages over $50$  independent walks.  Due to brevity requirements, only results for $T=5$ are given. As there is an inverse relationship between $\lambda$ and ruggedness, it can be observed that ruggedness decreases as the number of player gets larger. On the other hand, the number of players getting larger leads to a larger variety of $\lambda$, which means that the landscapes differ in ruggedness.  This effect is amplified by replacement restrictions, see \ref{fig:rugg}c-d. As this is not matched by a likewise increase in the number of maxima (compare to Fig. \ref{fig:modality}) additional topological features must be augmented by additional players.  Thus, it may be interesting to analyses the landscapes by further landscape measures, for instance information content. 

\section{Summary and conclusions} \label{sec:con}
Coevolutionary games have players that update their strategies as well as their networks of interaction. In this study, a
reinterpretation of coevolutionary games as dynamic fitness landscapes is proposed. The dynamic landscapes are based on three major components: (i) a description of strategy updating as a Moran process with definable  probabilities of strategy transitions, (ii) a formulation of network of interaction updating as instances of random regular graphs, and (iii) a linear relation between payoff and fitness.   Using these components, payoff--related fitness landscapes can be defined for each player.  It is further shown that coevolutionary  game dynamics can be expressed by a game landscape derived from these player--wise landscapes by including the strategy updating process. Moreover, different strategy updating processes, such as  death--birth (DB) or birth--death (BD) produce different game landscapes, which can be seen as strategy updating breaking the symmetry of the play--wise landscapes. In numerical experiments it has been demonstrated that landscape measures such as modality and ruggedness allow to differentiate between different game landscapes. 
So far the approach presented is a
technique for analyzing coevolutionary games by landscapes. However, by using the distribution of maxima in strategy space, it might be feasible to estimate fixation probabilities. 
Moreover, the approach is
not restricted to Moran processes as long as strategy transition probabilities can be derived, at least approximately.
Finally, networks updating is  currently modeled as a given sequence of random regular graphs, but should be understood as a transition process,
for instance by using
reproducing graphs~\cite{south10} as a tool to refine the description of transitions between adjacency matrices.


\begin{thebibliography}{1}

\bibitem{allen14} B. Allen and M. A. Nowak,  Games on graphs. {\em EMS Surv. Math. Sci.} {\bf 1}: 113--151,  2014.

\bibitem{bay10} M. Bayati,   J.~H. Kim, and A.  Saberi,  A sequential algorithm for generating random graphs.
 {\em Algorithmica} {\bf 58}: 860--910, 2010.

\bibitem{blitz11} J. Blitzstein and P.  Diaconis,  A sequential importance sampling algorithm for generating random graphs with prescribed degrees. {\em  Internet Mathematics} {\bf  6}: 489--522,  2011.

\bibitem{brede11}  M. Brede, The evolution of cooperation on correlated payoff landscapes. {\em Artificial Life} {\bf 17}: 365--373, 2011.

\bibitem{foster13} D.~V. Foster,   M.~M.~Rorick, T. Gesell, L.~M.~Feeney, and  J.~G. Foster,  Dynamic landscapes: a model of context and contingency in evolution. {\em J. Theor. Biol.} {\bf 334}: 162--172, 2013.

\bibitem{green12} G.~W. Greenwood and D.  Ashlock,  Evolutionary games and the study of cooperation: Why has so little progress been made? In: H. Abbass, D. Essam, R. Sarker (eds.), {\em Proc. IEEE Congress on Evolutionary Computation, IEEE CEC 2012},  IEEE Press, Piscataway, NJ,  1--8, 2012.

\bibitem{green14} G.~W. Greenwood and P. Avery, Does the Moran process hinder our understanding of cooperation in human populations? In:  G. Rudolph, M. Preuss (eds.), {\em Proc. IEEE Conference on Computational Intelligence and Games, IEEE CIG 2014},  IEEE Press, Piscataway, NJ, 1--6, 2014.



\bibitem{kauffm91} S.~A.~Kauffman and  S.~Johnsen, Coevolution to the edge of chaos: Coupled fitness landscapes, poised states, and
coevolutionary avalanches. {\em J. Theor. Biol.} {\bf149}:
467--505,  1991.

\bibitem{kim06}  J. H. Kim and   V. H. Vu,  Generating random regular graphs. In:  L. L. Larmore,
               M. X. Goemans (eds.), {\em Proc. ACM Symposium on Theory of Computing, STOC  2003},   ACM, New York,  213--222, 2003.

\bibitem{lieb05}  E.  Lieberman, C. Hauert, and  M.~A.   Nowak, Evolutionary dynamics on graphs.  {\em Nature} {\bf 433}: 312--316, 2005.

\bibitem{maysmit91} J. Maynard Smith,   {\em Evolution and the Theory of Games.} Cambridge University Press, Cambridge, 1991. 

\bibitem{nowak06}  M.~A. Nowak, {\em Evolutionary Dynamics: Exploring the Equations of Life.} Harvard University Press, Cambridge, MA,  2006. 

\bibitem{nowak93} M.~A. Nowak and R. M. May,  The spatial dilemmas of evolution. {\em Int. J. Bifurc. Chaos} {\bf 3}: 35--78, 1993. 


\bibitem{nowak04} M.~A. Nowak and K. Sigmund, Evolutionary dynamics of biological games. {\em Science} {\bf 303}: 793--799,  2004. 



\bibitem{ohts07} H. Ohtsuki, J.~M.~Pacheco, and  M.~A. Nowak,   Evolutionary graph theory: Breaking the symmetry between interaction and replacement. {\em J. Theor. Biol.} {\bf 246}: 681--694, 2007.

\bibitem{pach06} J.~M.~Pacheco, A. Traulsen, and M.~A. Nowak, Coevolution of strategy and structure in complex networks with dynamical linking. {\em Phys. Rev. Lett.} {\bf 97}: 258103, 2006. 


\bibitem{patt15} K. Pattni, M. Broom, L. Silvers, and  J. Rychtar. Evolutionary graph theory revisited: When is an evolutionary process equivalent to the Moran process? {\em Proc. Roy. Soc.} {\bf A471}: 20150334, 2015.

\bibitem{perc10} M. Perc and A. Szolnoki,  Coevolutionary games--A mini review. {\em BioSystems} {\bf 99}: 109--125, 2010.




\bibitem {rich08} H.~Richter,
Coupled map lattices as spatio--temporal fitness functions:
Landscape measures and evolutionary optimization. {\em Physica D}
{\bf 237}:  167--186, 2008.

\bibitem{rich14a} H. Richter, Fitness landscapes that depend on time. In: H. Richter, A. P. Engelbrecht (eds.), {\em Recent Advances in the Theory and Application of Fitness Landscapes}, Springer--Verlag, Berlin, 265--299, 2014.

\bibitem{rich14b} H.~Richter,
 Codynamic fitness landscapes of coevolutionary minimal substrates. In:  C. A. Coello Coello (ed.), {\em Proc. IEEE Congress on Evolutionary Computation, IEEE CEC 2014},  IEEE Press, Piscataway, NJ,  2692--2699, 2014.

\bibitem{rich15} H.~Richter,  Coevolutionary intransitivity in games: A landscape ana\-lysis. In:  A.~M.~Mora, G.~Squillero (eds.), {\em Applications of Evolutionary Computation - EvoApplications 2015},  Springer--Verlag, Berlin, 869--881, 2015.

\bibitem{richengel14}  H.~Richter and  A.~P. Engelbrecht, {\em Recent Advances in the Theory and Application of Fitness Landscapes.}  Springer--Verlag, Berlin, 2014.  

\bibitem{sha12} P. Shakarian, P. Roos, and A. Johnson, A review of evolutionary graph theory with applications to game theory. {\em  BioSystems} {\bf 107}: 66--80, 2012.



\bibitem{slon15} N. Sloane, The on--line encyclopedia of integer sequence.
 https://oeis.org (retrieved October, 29, 2015).

\bibitem{south10} R. Southwell and C. Cannings, Some models of reproducing graphs. I. Pure reproduction. {\em Appl. Math.} {\bf 1}: 137--145,  2010.

\bibitem{stad03} P.~F. Stadler and  C.~R. Stephens,    Landscapes and
effective fitness. {\em Comm. Theor. Biol.} {\bf  8}: 389--431, 2003. 



\bibitem{szabo07}  G. Szabo and  G.~Fath,   Evolutionary games on graphs. {\em Phys. Rep.} {\bf 446}: 97--216, 2007.

\bibitem{tani07} J. Tanimoto, Dilemma solving by the coevolution of networks and strategy in a 2 $\times$ 2 game. {\em Phys. Rev.} {\bf E76}: 021126-1-7,   2007.





\bibitem{whig08}  P.~A. Whigham and G. Dick,  Evolutionary dynamics for the spatial Moran process. {\em Genet. Programm. Evol. Machine} {\bf 9}: 157--170, 2008.

\bibitem{worm99}  N. C. Wormald,   Models of random regular graphs.  In: J.~D.
Lamb,  D.~A. Preece (eds.),
{\em Surveys in Combinatorics, 1999},   Cambridge University Press, Cambridge, 239--298, 1999.


\end{thebibliography}
\end{document}